\documentclass[journal abbreviation, final]{copernicus}
\usepackage{makecell}
\usepackage{subfigure}
\usepackage[nice]{nicefrac}
\usepackage{multirow}
\usepackage{float}
\usepackage{tabularx}
\usepackage{xcolor}
\usepackage{url}
\usepackage{comment}
\usepackage{dblfloatfix}
\usepackage{booktabs}
\usepackage{tabularx}
\usepackage{array}
\usepackage{amsmath}
\usepackage{array}
\usepackage{placeins}
\usepackage{xcolor}
\usepackage{siunitx}[=v2]

\begin{document}

\title{Surrogate-Assisted Framework for SI-Compliant Interconnect Design Optimization Using the Earth Mover's Distance}
\Author[1]{Emre}{Ecik}
\Author[2]{Werner}{John}
\Author[1]{Julian}{Withöft}
\Author[3]{Ralf}{Brüning}
\Author[1]{Jürgen}{Götze}

\affil[1]{Information Processing Lab, TU Dortmund University, 44227 Dortmund, Germany}
\affil[2]{Pyramide2525/TU Dortmund University, 33100 Paderborn, Germany}
\affil[3]{EMC Technology Center Paderborn, Zuken GmbH, 33100 Paderborn, Germany}

\correspondence{Emre Ecik (emre.ecik@tu-dortmund.de)\\\vspace{2mm}\small\textit{This manuscript has been submitted to Advances in Radio Science for review (2026).}\\\vspace*{-0.5cm}}

\runningtitle{Surrogate-Assisted SI-Compliant Interconnect Optimization Using the Earth Mover's Distance}

\runningauthor{E. Ecik et al.}

\received{}
\pubdiscuss{}
\revised{}
\accepted{}
\published{}

\firstpage{1}

\maketitle
\nolinenumbers

\begin{abstract}
This work presents a deterministic, machine-assisted framework for SI-compliant PCB design based on the Earth Mover’s Distance (EMD). In contrast to conventional surrogate-based optimization methods that rely on iterative black-box search procedures, the proposed approach follows an interpretable, sequential evaluation strategy. Neural surrogate models are first used to efficiently predict waveform describing features from topology-dependent design parameters. A decision tree then acts as a physically motivated quality gate that identifies SI-compliant waveforms according to predefined SI criteria. Within the resulting valid solution space, the Earth Mover’s Distance is employed as a similarity metric to rank candidate designs according to their proximity to an ideal reference signal. This enables not only the deterministic identification of admissible parameter regions but also a transparent prioritization of physically superior solutions without inverse modeling or stochastic search procedures. The methodology is demonstrated using a large-scale set of simulated DDR3 fly-by waveforms. By combining surrogate prediction, interpretable classification, and EMD-based waveform evaluation, the framework provides an explainable and computationally efficient alternative to conventional optimization strategies for supporting PCB development with AI-based methods.
\end{abstract}

\introduction  
Modern EDA tools increasingly integrate methods of machine learning (ML) to address the complex challenges of PCB design more efficiently \citep{zuken}. In this context, ML techniques support designers in obtaining reliable results more quickly by using surrogate models to approximate, for example, eye diagrams \citep{ma} and S parameters \citep{li}. In this way, time consuming field solvers can be replaced by fast predictions, which directly improves both time to market (TTM) and time to volume (TTV). The combination of surrogate models with optimization algorithms has been used in the literature to propose solution strategies for efficient and developer oriented PCB design that systematically accounts for the stringent requirements of signal integrity \citep{torun,zhang,medico}. In these approaches, surrogate models are typically employed as fast substitutes to iteratively search for optimal solutions within a large parameter space. Such classical surrogate based optimizers primarily focus on the efficient identification of SI compliant parameters. Since the emphasis is placed on convergence, they often provide little insight into the underlying decisions, which can represent a black box, particularly for less experienced designers. Moreover, a fixed model is usually employed, and the iterative search in the design space proceeds without refinement of the prediction model during the process. In addition to these approaches, the literature also reports two stage procedures for finding optimal solutions that follow a different methodological principle. These are domain independent optimization frameworks that have been presented as methodological concepts, for example in aerodynamics or materials science \citep{alexandrov, pourkamali}. They employ refinement oriented ideas to successively improve model quality by comparing different levels of accuracy or by systematically reducing statistical uncertainty.
The approaches mentioned above differ fundamentally from the focus proposed in this work. The emphasis of the presented results lies on the physical assessment and differentiation of design solutions that have already been identified as valid, with the aim of supporting designers in PCB development. Iterative black box optimizations are completely avoided and surrogate models are instead used exclusively as reliable physical data sources for a structured analysis of the design space. The method is deterministic and does not require any iterative search, which leads to higher computational efficiency and improved interpretability compared to approaches such as GA or BO. Rather than searching for a single optimal point, the method relies on decision trees to evaluate signal integrity within a design space captured by comprehensive parameter space sampling \citep{ecik2023,ecik2024}. Furthermore, no model is refined or inverted; instead, only physically validated solutions are evaluated. The entire process thus remains a forward directed approach, in which parameters are mapped unambiguously from the signal waveforms. This avoids the non uniqueness issues typical of inverse methods, where a single observed output can correspond to multiple distinct input configurations. While inverse approaches such as \cite{pourkamali} must rely on statistical uncertainty models due to this mathematical underdetermination, this issue is structurally excluded in the framework presented here. If different parameter combinations lead to identical or similarly good SI signals, they are simply treated as valid candidate design solutions within the model. The evaluation is always carried out deterministically based on physical criteria. This results in a transparent, explainable, and fully forward oriented optimization procedure that operates without inverse modeling steps and without statistical uncertainty estimates. The focus is therefore on interpretability in the sense of a white box model, whereby the explainability primarily concerns the downstream logic. First, a physically validated subset of SI compliant designs is identified and then partitioned into logically interpretable regions by the tree structure. The methodological extension introduced here through the use of the Earth Mover’s Distance (EMD) represents a significant advancement over the purely statistical analysis based on probability mass functions (PMF) in \cite{ecik2024}. While the PMF based approach aims to identify robust design regions by statistically evaluating valid parameter combinations, the EMD enables an additional and substantially more precise differentiation of the results by means of a physically motivated metric.
Originally introduced by \cite{rubner} for computer vision, the EMD (also known as the Wasserstein metric) is used here to assess signal integrity by measuring the distance between a waveform and a reference. The waveform is interpreted as a normalized amplitude distribution, allowing the EMD to quantify the amount of work required to transform a given signal shape into the ideal reference shape. Since the EMD establishes a metrically grounded ranking among the SI compliant solutions, it enables a clear qualitative assessment of the waveforms themselves. The focus thus shifts from merely identifying safe design regions to the deterministic selection of the physically optimal solution within a transparently defined SI compliant domain. This combines the advantages of full design space coverage through parameter space sampling with the transparency of a decision tree and the physical precision of the EMD metric, thereby forming a sequential framework for design optimization (see Figure \ref{fig:framework}).
Conceptually, the implementation of the decision tree is inspired by classical expert systems. These systems, which shaped AI research in the 1960s and 1970s - prominently represented by pioneering projects such as DENDRAL \citep{dendral} or MYCIN \citep{mycin} - are based on explicitly formulated if-then rules. In the present work, the definition of specific SI criteria is first used to establish a consistent labeling structure of the dataset as a physically grounded ground truth. To overcome the limitations of purely hand-coded rule structures, this approach is then transferred into a generalizing expert model. While a classical expert system relies on rigidly programmed rules, the decision tree derives them automatically from examples and identifies physical patterns in a data driven manner. This transition from a controlled rule base to a physically motivated generalization enables the correct classification of new, previously unseen signal waveforms. The decision tree thus functions as a model based quality gate that ensures deterministic and explainable validation.
The following section establishes the methodological foundations required for the proposed framework (see Figure \ref{fig:framework}). Subsequently, it details the sequential, physically inspired workflow designed to identify optimal SI-compliant signals and their corresponding design parameters.

\section{Methodology}
\begin{figure}[h]
\centering
    \includegraphics[width=0.48\textwidth]{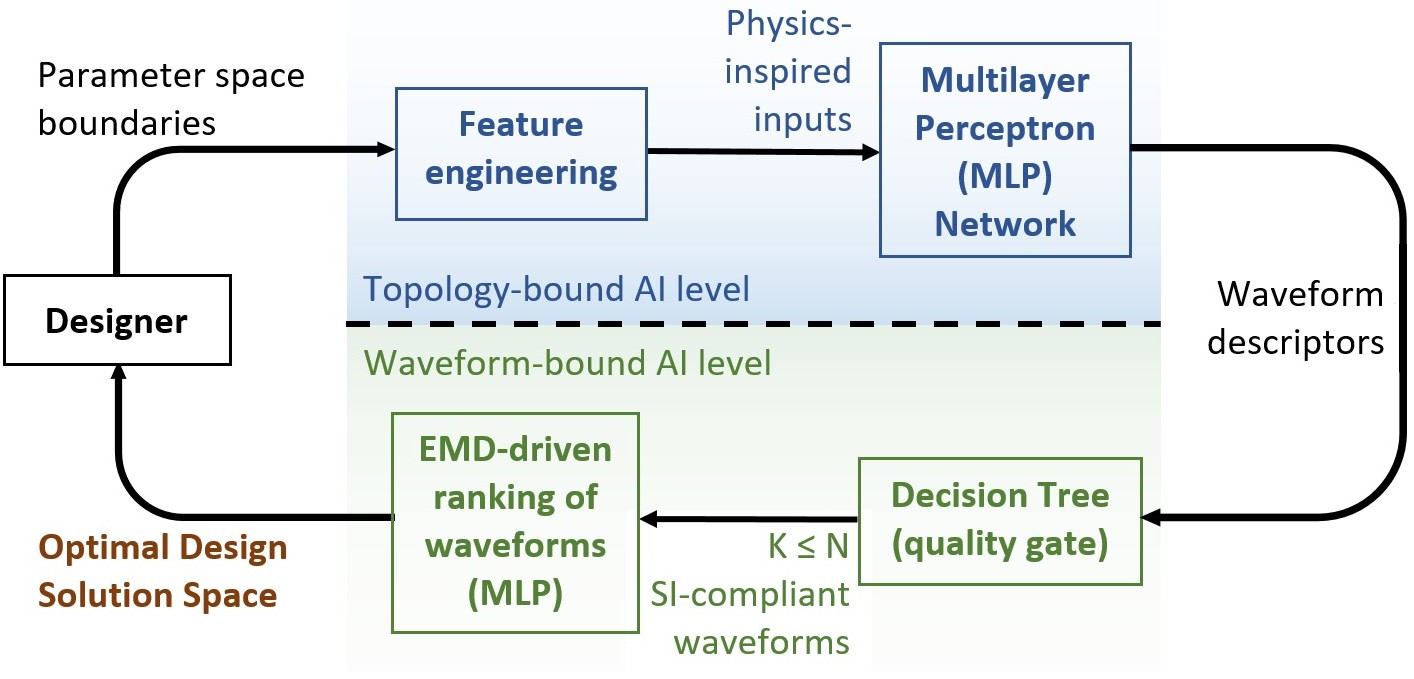}
    \caption{EMD-Enhanced Optimization Framework for SI-Compliant Parameter Space Exploration}
    \label{fig:framework}
\end{figure}

The methodology presented in this work follows a sequential logic (see Fig. \ref{fig:framework}) and integrates neural surrogate models for prediction, decision trees for interpretability, and the EMD metric for a physically grounded evaluation into a deterministic optimization framework.
At the beginning, the designer defines the parameter space boundaries, thereby specifying both the region to be explored and the associated design constraints. From these parameter combinations, a physics inspired feature engineering script first generates domain specific input features. This step is essential, as it introduces domain knowledge in a targeted manner. The small number of circuit parameters is transformed into a larger set of physically meaningful input quantities, providing the subsequent multilayer perceptron (MLP) network with crucial learning support. The MLP then generates the waveform descriptors, which precisely characterize the properties of the underlying signals and serve as input for the next stage. A detailed description of the data generation phase is provided in Section \ref{dataGen}. After the waveform descriptors have been computed, the decision tree - acting as a quality gate - selects the SI compliant signals from the proposed parameter space. Afterward, the EMD values are computed via a surrogate model and compared with a reference waveform that represents an ideal SI compliant signal (maximum possible slew rate with minimal overshoot). By selecting the waveforms with the lowest EMD values - i.e., those most similar to the reference signal - the designer obtains a ranked list of optimal design parameters within the originally defined boundaries. It is important to note that the feature-engineering module and the MLP operate on parameter- and topology-bounded level, while the decision tree and the EMD surrogate are applied exclusively at the waveform level, where the actual signal evaluation takes place. 

Furthermore, this work adopts the assumption that the methodology focuses specifically on the signal integrity pipeline and prioritizes the optimization of signal morphology. To ensure a clear assessment of algorithmic performance in evaluating the signal shape, temporal parameters (e.g., signal propagation delays and bus skew) were deliberately treated as independent variables and excluded from the optimization framework. This separation of concerns aims to preserve a clean, detectable signal edge as a fundamental physical prerequisite before higher-level timing synchronization or hardware-based compensation mechanisms (e.g., DDR3 write leveling) can operate effectively.

\subsection{Decision Trees}\label{decisionTree}
The decision tree used in this work is based on the CART algorithm \citep{brei} as implemented in the MATLAB environment \citep{matlab}. The structural hyperparameters of the tree were determined using Bayesian optimization and subsequently adopted as the starting point for the final training. This procedure is based on the pioneering work of \cite{kushner} and \cite{mockus}, which aims to efficiently identify the global optimum of an unknown objective function. In its modern form, the framework follows the principle of Efficient Global Optimization \citep[EGO;][]{jones}.

Although the CART method is classically associated with the Gini index \citep{gini} as its traditional impurity criterion, the hyperparameter optimization conducted in this work indicated a data-dependent selection of the split criterion. For smaller training datasets, the Gini-based criterion was selected, whereas for substantially larger datasets, entropy \citep{shannon} provided superior separation performance and was therefore chosen during optimization. Consequently, the implemented model retains the structural framework of CART while employing entropy-based split evaluation for larger training sets, consistent with the formulation of the C4.5 decision tree algorithm \citep{quin}. This results in a data-driven hybrid decision tree formulation that adapts the impurity criterion to the respective training scenario.

In addition, a cost matrix $C$ was used during the training of the decision tree. Since SI-compliant waveforms constituted only a small fraction of the dataset, false negatives (valid designs incorrectly classified as non–SI-compliant) were weighted more strongly than false positives (non–SI-compliant designs incorrectly classified as valid) via the parameter $ratio$:

\begin{equation}
    C = \begin{pmatrix} 
    0 & 1 \\ 
    ratio & 0
    \end{pmatrix}.
\label{costmat}
\end{equation}

While the structural tree parameters were determined by Bayesian optimization, the parameter $ratio$ was set in a subsequent empirical analysis. For this purpose, the misclassification cost weighting was gradually adjusted across multiple training runs to counteract the underrepresentation of the SI-compliant class in the learning algorithm. The objective was to reduce false negatives in a targeted manner without causing a disproportionate increase in false positives.

\subsection{MLP and Residual Networks}
Multi-Layer Perceptrons (MLPs) are fully connected feedforward networks that model nonlinear relationships through a sequence of linear transformations followed by activation functions. A fully connected layer maps an input vector $x$ by computing weighted sums of the inputs and applying an activation function $\phi$ to obtain the output vector $y$.

For each component $y_k$ of the output vector, the following holds \citep{matlab, bishop}:
\begin{equation}
    y_k = \phi \left( \sum_{i} W_{k,i} x_i + b_k \right).
\end{equation}
Here, $W_{k,i}$ denote the weights, $b_k$ the bias term, and $x_i$ the components of the input vector.

This work employs the Gaussian Error Linear Unit \citep[GELU;][]{matlab, hendrycks} as the activation function. GELU is a probabilistic activation defined as:
\begin{equation}
    \text{GELU}(x) = x \cdot \Phi(x) = \frac{x}{2} \left( 1 + \text{erf} \left( \frac{x}{\sqrt{2}} \right) \right),
\end{equation}
where $\Phi(x)$ represents the cumulative distribution function of the standard normal distribution. By avoiding hard thresholding, the GELU activation provides smoother gradients and supports stable convergence during training.

To improve the trainability of deeper models, this work applies a residual learning approach as introduced by \cite{he}. Instead of learning a direct mapping, the network learns a residual function $F(x)$. The resulting output is given by:
\begin{equation}
    y_k = F_k(x) + x_k.
\end{equation}
This structure is implemented through skip connections that forward the input $x$ unchanged to deeper layers, stabilizing gradient flow and enabling efficient training of deep architectures.

Layer Normalization \citep{matlab, ba} is used to further stabilize the training dynamics. For each sample, the mean $\mu$ and variance $\sigma^2$ are computed across all feature dimensions, and the normalized activations are obtained as:
\begin{equation}
    \hat{x}_i = \frac{x_i - \mu}{\sqrt{\sigma^2 + \epsilon}}.
\end{equation}
This yields consistent activation statistics independent of batch size and improves the robustness of the training process.

\subsection{One-Dimensional Wasserstein Distance (Earth Movers Distance) and its Discrete Formulation}
The goal of the Wasserstein distance is to quantify the similarity between two probability distributions. The concept originates from the theory of optimal transport introduced by \citet{kantor}, and was later formulated as a metric on probability measures by \citet{vaser}. In many applications, the Wasserstein distance is also referred to as the Earth Movers Distance (EMD), a term that became widely established in image processing through the work of \citet{rubner}.

Since the general form of the optimal transport problem is difficult to solve, \citet{vallender} provided an explicit proof and a closed-form expression for the one-dimensional special case. In the one-dimensional setting, the $W_1$ distance simplifies considerably. The similarity between two distributions $P$ and $Q$ on the real line with finite first moments can be expressed as the integral of the absolute differences of their cumulative distribution functions (CDFs):
\begin{equation}
W_1(P,Q) = \int_{-\infty}^{\infty} \lvert F_P(x) - F_Q(x) \rvert \, dx.
\end{equation}

Vallender showed that the one-dimensional Wasserstein distance can equivalently be expressed in terms of quantile functions.:
\begin{equation}
\int_{0}^{1} \lvert F_P^{-1}(u) - F_Q^{-1}(u) \rvert \, du
=
\int_{-\infty}^{\infty} \lvert F_P(x) - F_Q(x) \rvert \, dx .
\end{equation}

This duality is particularly important in the discrete case, where two vectors are to be compared. In this setting, the empirical quantile functions are obtained by sorting the samples of the two vectors $x$ and $y$. As a result, the W1 distance in the discrete case can be written as:
\begin{equation}
\label{W1eq}
W_1(x,y) = \frac{1}{L} \sum_{i=1}^{L} \lvert x_{(i)} - y_{(i)} \rvert ,
\end{equation}
where $x_{(i)}$ and $y_{(i)}$ denote the sorted values of the two vectors. L denotes the common length of the two vectors, corresponding to empirical distributions with uniform weights.

\subsection{Genetic Algorithm}
In analogy to previous results in \cite{ecik2024}, the genetic algorithm (GA) was used in this work as a reference method to enable a comparison with other approaches from the literature \citep{zhang}. The genetic algorithm was developed by \cite{holland} and is today classified as a subfield of evolutionary algorithms. This optimization method is based on the principle of natural evolution, starting with a random population of candidate solutions. The quality of each individual is evaluated using a fitness function, defined as a minimization measure that indicates how well a candidate solution satisfies the optimization objective. Through the genetic operators crossover (recombination of features from two individuals to generate offspring) and mutation (random modification of individual components), new populations are generated over multiple generations. The goal is to gradually identify better solutions to the given optimization problem.

In this work, the fitness function of the genetic algorithm is defined by the Earth Mover’s Distance. This approach ensures comparability with the framework introduced in Figure \ref{fig:framework} by establishing a common evaluation basis grounded in the EMD. In contrast to the decision tree approach presented here, the genetic algorithm represents a population-based optimization procedure. The decision tree approach relies on a deterministic selection mechanism in which suitable candidate solutions are first identified and those with minimal EMD are then selected directly through a global selection process rather than through an iterative search.

\subsection{Evaluation Metrics}
For evaluating the model performance, different metrics were employed depending on the underlying learning task. Since the decision tree addresses a binary classification problem, classification-specific metrics such as sensitivity (recall), specificity, and the Matthews correlation coefficient (MCC) were applied. These metrics are based on the entries of the confusion matrix and are therefore well suited to quantify the predictive quality of binary class assignments. In particular, the MCC is advantageous for strongly imbalanced datasets, as it incorporates true positives (TP), true negatives (TN), false positives (FP), and false negatives (FN) into a single balanced performance indicator.

In contrast, the surrogate models represent regression tasks, as they predict continuous target values rather than discrete class labels. Consequently, classification metrics based on confusion matrix statistics are not applicable in this context. Instead, the normalized root mean squared error (NRMSE) and the Pearson correlation coefficient (PCC) $r$ were used for performance evaluation. The NRMSE quantifies the absolute prediction deviation in a normalized and dimensionless form, enabling scale-independent comparisons across different datasets, model variants, and target variables. To ensure a data-leakage-free error assessment, the normalization parameters for the NRMSE calculations were determined exclusively from the respective training datasets and subsequently applied to all corresponding test datasets. In addition, the Pearson correlation coefficient evaluates the linear agreement between predicted and actual target values, thereby providing a complementary assessment of trend consistency.

The mathematical definitions of all evaluation metrics used are summarized in Table \ref{mathMetricsTable} in the Appendix \ref{mathMetrics}.

\section{Data Generation and Processing Pipeline}\label{dataGen}
\begin{figure}[h]
\centering
    \includegraphics[width=0.48\textwidth]{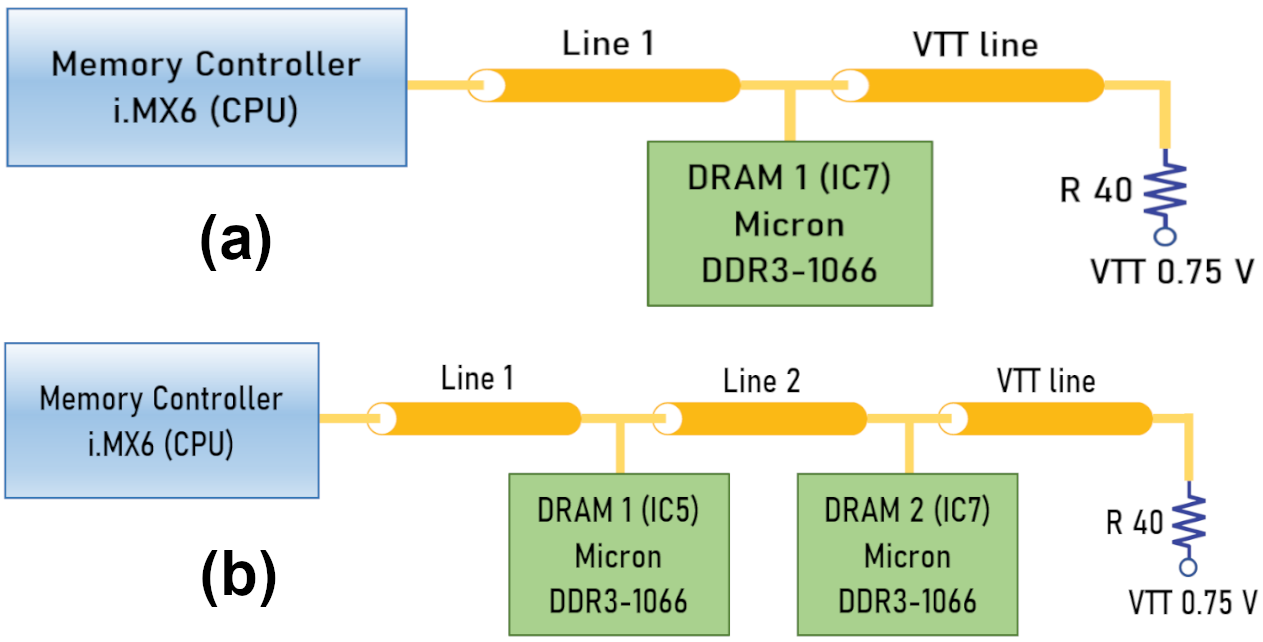}
    \caption{Utilized DDR3 fly-by networks: (a) Simulation with single DRAM module; (b) Simulation with two DRAM modules}
    \label{fig:netze}
\end{figure}

Matlab \citep{matlab} and the eCadstar SI simulation tool \citep{zuken} were used to generate a very large dataset of simulations in batch mode. The investigated single-ended DDR3 network strcutures are illustrated in Figure \ref{fig:netze}. 

In total, 125,000 simulations were performed for a DDR3 fly-by networks with one DRAM module and 492,075 simulations were carried out for a configuration with two modules. Since each DRAM represents an individual observation point, this resulted in a total of 125,000 and 984,150 extracted waveforms, respectively. The specific parameter variations used to generate this dataset, including trace lengths and VTT termination resistance values for a characteristic impedance of $Z=50\,\Omega$, are detailed in Table \ref{tab:sweeps}.

\begin{table}[h]
\centering
\caption{Parameter variations for the DDR3 fly-by networks}
\label{tab:sweeps}
\begin{tabular}{lcc}
\toprule
 & One DRAM & Two DRAMs \\
\midrule
Line 1 [mm]    & 1 to 491 ($\Delta$ 10) & 1 to 391 ($\Delta$ 15) \\
Line 2 [mm]    & N/A                    & 1 to 391 ($\Delta$ 15) \\
VTT line [mm]  & 1 to 491 ($\Delta$ 10) & 1 to 391 ($\Delta$ 15) \\
R 40 [$\Omega$] & 1 to 491 ($\Delta$ 10) & 1 to 481 ($\Delta$ 20) \\
\midrule 
No. of simulations & 124,999 & 492,075 \\
No. of waveforms   & 125,000 & 984,150 \\ 
\bottomrule
\end{tabular}
\end{table}

As in previous work \citep{ecik2023,ecik2024}, a stimulus frequency of 133 MHz was used and the simulations were performed using IBIS models for the memory controller and DRAM chips. 

\begin{figure}[h]
\centering
    \includegraphics[width=0.48\textwidth]{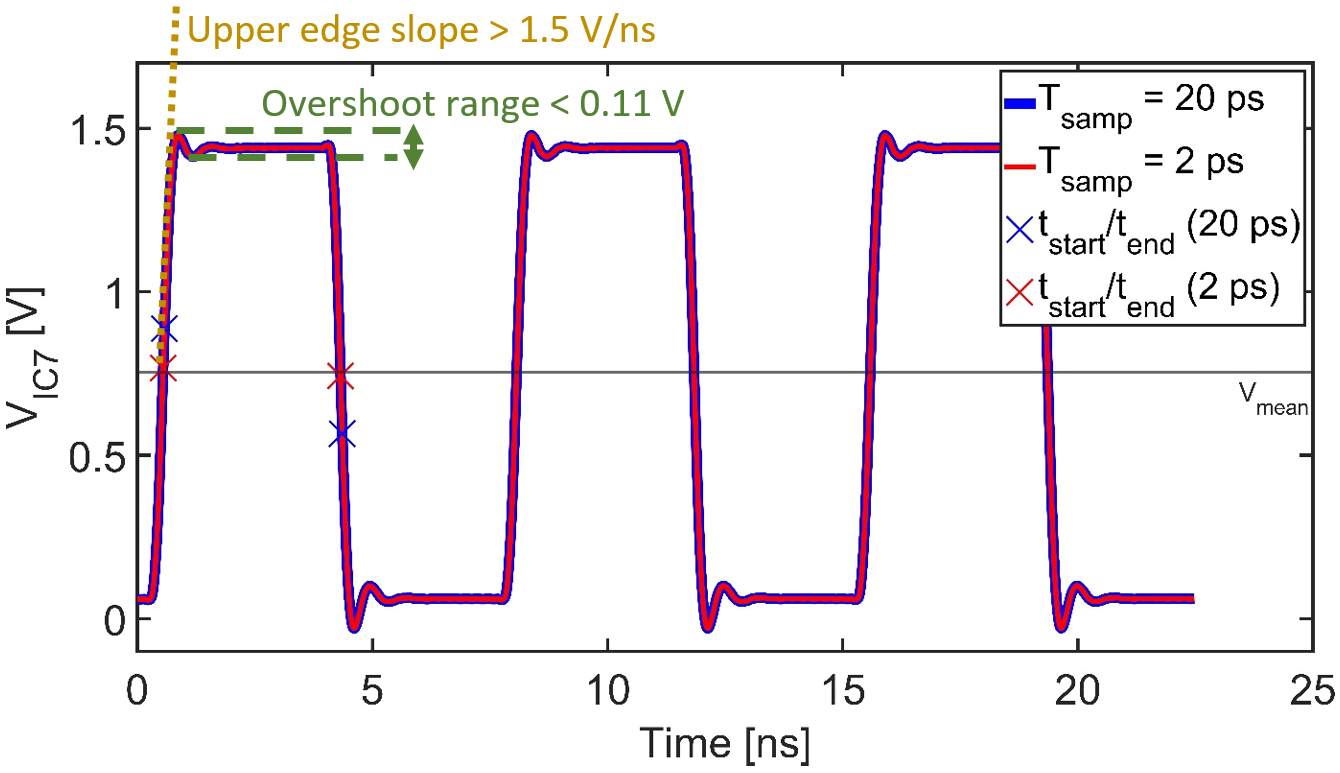}
    \caption{Comparison of an interpolated and original SI-compliant waveform (including overshoot range and upper edge slope criteria) for mid-mevel intersection determination}
    \label{fig:siKonf}
\end{figure}

After completing the simulations, the first period was consistently removed from all signals to avoid transient effects, leaving only three signal periods for evaluation. To reduce the simulation time given the large number of runs, the sampling interval during the simulations was set to 20~ps. During signal processing, it was then linearly interpolated to 2~ps, which improved the accuracy of the threshold evaluations in the subsequent steps. As shown in Figure \ref{fig:siKonf}, this type of interpolation noticeably improved the determination of the first two intersections of the waveform with its mid-level value for an SI-compliant signal. The red crosses indicate the intersection points obtained after interpolation, while the blue crosses correspond to the original waveform sampled at 20~ps. These intersection points, obtained from the interpolated (red) waveforms, define the start and end of the extracted signal segment and served as the basis for the subsequent labeling and extraction of waveform descriptors.

Since a clear distinction between distortion-free and degraded signals was required for training the model, the data underwent a labeling process. To generate a reliable ground truth, a rule-based script was first applied to categorize the simulation data according to defined SI criteria. The signal quality assessment was based on key metrics, including: (a) compliance with overshoot limits (< 0.11 V), (b) adherence to a minimum upper edge slope (> 1.5 V/ns), and (c) preservation of the logic levels ($V_{inl(max)}$/$V_{inh(min)}$) specified in the IBIS receiver model and in \citet{jedec}. To convert this rigid logic into a generalized and robust methodology, the signals were subsequently transformed into physically meaningful waveform descriptors for training the decision-tree. This transformation from raw signals to waveform descriptors forms the essential link that enables the generalization capabilities of a decision tree while preserving physical interpretability. In addition, the same waveform descriptor set is employed for surrogate models, allowing for fast predictions without rerunning computationally expensive simulations during inference. The extracted waveform descriptors are categorized into time-domain and frequency-domain representations. A detailed overview of the descriptors used to transform the signals into the physical domain is provided in Table \ref{tab:featTab} in the Appendix \ref{appa} section.

For the final step of the data processing pipeline, the Wasserstein distances of the waveforms were computed according to Equation~\ref{W1eq}. The evaluation is performed on the extracted signal segments, as described previously. To ensure comparability of the Wasserstein distances across all signals, the amplitudes were scaled using min-max normalization.

\begin{figure}[h]
\centering
    \includegraphics[width=0.48\textwidth]{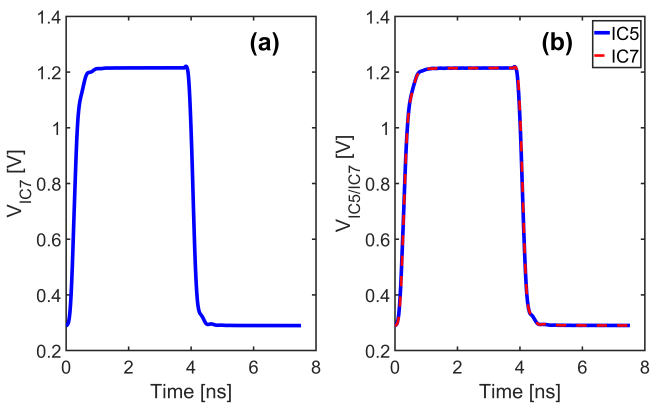}
    \caption{Reference waveform used for the calculation of the EMD values across network configurations: (a) One DRAM module (see Figure \ref{fig:netze} (a)); (b) two DRAM modules (see Figure \ref{fig:netze} (b))}
    \label{referenzSig}
\end{figure}

For the comparison, an ideal rectangular reference signal was assumed. The reference signals were generated using the networks shown in Figure \ref{fig:netze}. The line lengths were fixed at 1 mm, and the resistor in the TTL branch was set to 50 ohms to match the characteristic impedance Z of the networks. Figure~\ref{referenzSig} shows these reference signals for the configurations with one and two DDR3 modules. The signal shape was intentionally designed to provide adequate edge steepness while ensuring an overshoot-free waveform.

\begin{figure}[h]
\centering
    \includegraphics[width=0.48\textwidth]{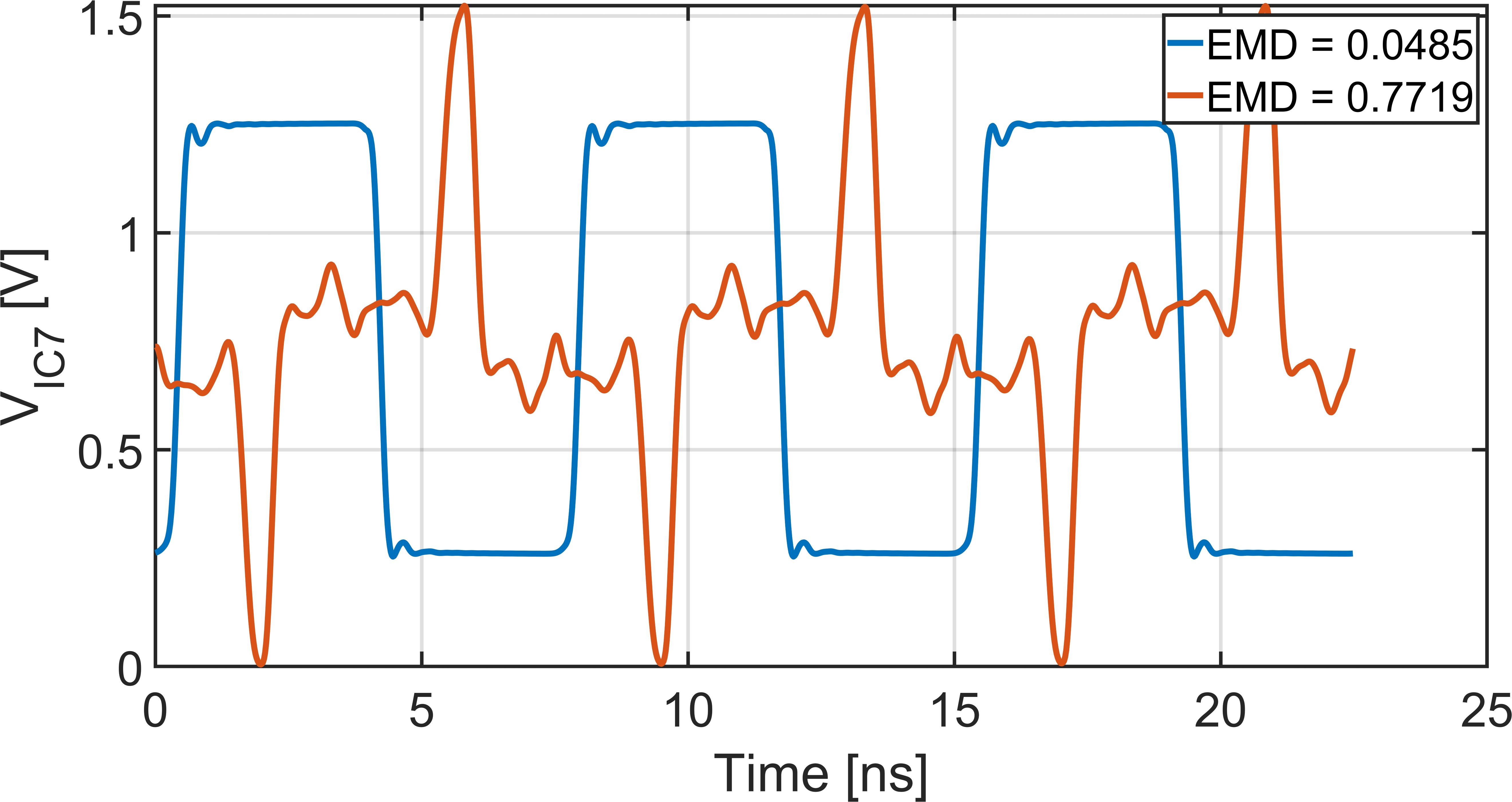}
    \caption{Comparison between an SI-compliant waveform with low W1 distance and a non-SI-compliant waveform with higher W1 distance}
    \label{fig:W1dist}
\end{figure}

Figure~\ref{fig:W1dist} shows an exemplary comparison between an SI-compliant and a non-SI-compliant waveform. It can be observed that the non-SI-compliant waveform exhibits a significantly larger $W_1$ distance compared to the nearly rectangular reference signal.

It is important to emphasize that the Wasserstein-based metric does not perform any intrinsic SI verification. Instead, it solely quantifies the similarity to the ideal rectangular shape. The actual SI classification is carried out by the decision tree in a previous step.

Both the training of the decision tree and the training of the surrogate models used to accelerate the predictions are described in the following section.

\section{Implementation of a Physics-Inspired Multi-Model Pipeline}
In addition to training the decision tree, further models are required to support the developer and to avoid costly simulations and complex computations - the so called surrogate models. In this work, a multilayer perceptron network with a residual block and layer normalization serves as the starting point. The predictions of this stage act as inputs to the subsequent models, resulting in a sequential modeling approach. Consequently, these predictions are essential and must be obtained with high accuracy.
Regarding the input for the multilayer perceptron network, the parameter combinations of the two networks described in Table \ref{tab:sweeps} were initially used. However, it became evident during implementation that experiments based solely on the raw parameters - both for the network with three input variables and for the model with four input variables (see Figure \ref{fig:netze}) - were insufficient. The available information was too limited to reliably predict the highly complex target quantities (see Table \ref{tab:featTab} in the Appendix \ref{appa}).
The waveform descriptors to be approximated arise from nonlinear transformations of the signal, such as derivatives, windowing operations, energy and entropy measures, and spectral analyses. As a result, they exhibit substantially higher structural complexity.
To address this issue, additional features were constructed in the sense of physics inspired feature engineering. These features reflect both physical relationships and the hierarchical structure of the data generation process. They include normalized base quantities, geometric relations between trace lengths, physically motivated characteristics such as reflection coefficient and conductance, harmonic Fourier features, and interaction terms.
Through this extension, the model receives preprocessed, structured, and physically interpretable information. This improves convergence and significantly enhances generalization capability. Specifically, the input space was expanded to a dimension of $d=62$ for the configuration with a single DDR3 module (see Fig. \ref{fig:netze} (a)) and $d=92$ for the dual-module setup (see Fig. \ref{fig:netze} (b)). A complete overview of all constructed features is provided in Table \ref{tab:engineered_features} in the Appendix~\ref{appb}.

Based on this foundation, the surrogate models described in the following section were subsequently implemented.

\subsection{Implementation of a Multilayer Perceptron Network for Waveform Descriptor Prediction}\label{MLPDescr}
\begin{figure*}[h]
\centering
    \includegraphics[width=0.8\textwidth]{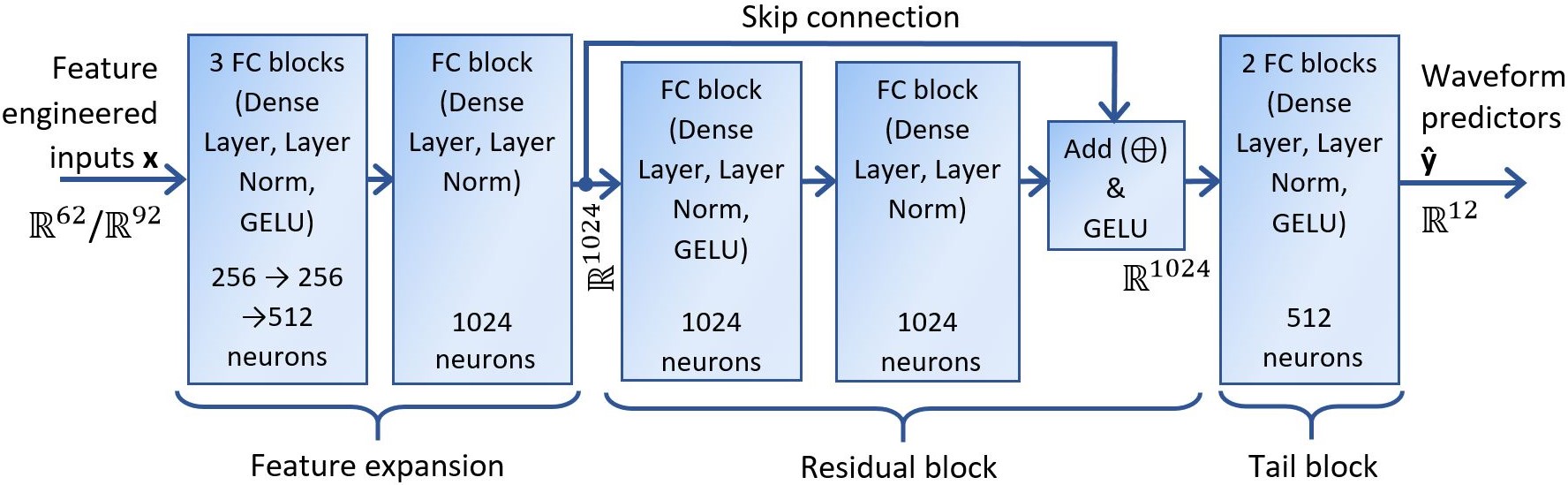}
    \caption{Architecture of the proposed surrogate neural network for waveform descriptor prediction}
    \label{fig:confSurrFeat}
\end{figure*}

The implemented model was designed as a residual deep multilayer perceptron (MLP) developed specifically for stable and high-precision feature representation. Its structural layout is divided into three main components. First, the data pass through a feature-expansion network consisting of four fully connected (FC) layers. This stage projects the input features stepwise through 256, 256, and 512 neurons into a 1024-dimensional representation space. Each layer is stabilized by layer normalization and uses the GELU activation function. The core of the architecture is a residual block with a skip connection. This block splits the 1024-dimensional signal into a shortcut path that preserves the identity mapping and a main path that applies the nonlinear transformation. Their subsequent recombination by addition reduces gradient-related issues and increases the model’s capacity, enabling the learning of fine-grained nonlinear corrections. The final component is the tail section. This output stage maps the refined representations from the residual block to the target parameters through two additional specialized layers, each with 512 neurons. The complete structure of the surrogate network is shown in Figure \ref{fig:confSurrFeat}.

The choice of a deep residual architecture is motivated by the high structural complexity of the underlying waveform descriptors. These descriptors result from strongly nonlinear transformations of the signal and exhibit an elevated level of representational complexity. In addition, a large training dataset of approximately 0.8⋅492,075 samples is available, providing sufficient data complexity to reliably train a high-capacity model.
Smaller network architectures tested in preliminary experiments showed insufficient model capacity and tended to represent the nonlinear relationships within the feature space only incompletely. The selected residual deep MLP therefore enables a significantly improved approximation of the underlying functional structure while maintaining stable optimizability. This component plays a central role within the overall system, as its outputs serve as high-precision and stable representations for downstream models in the pipeline. As a result, the network functions not only as a regression model but also as a deterministic feature generator within a multi-stage modeling framework. The quality of these representations directly affects the performance of the subsequent classification models. The choice of a deep architecture is therefore further motivated by the requirement to minimize error propagation within the model cascade. 

Training for both models was performed over 600 epochs. To achieve stable convergence together with strong generalization, a unified one-cycle learning-rate schedule \citep{smith} with annealing was applied. During an initial warm-up phase, the learning rate increased linearly from a starting value of 1e−4 to a maximum of 1e−3, followed by a controlled cosine-based decay to a final minimum of 1e−5. The computational workload was executed on a high-performance hardware system consisting of an Intel Xeon Platinum 8268 CPU and an NVIDIA RTX A6000 GPU. The large GPU memory capacity of 48 GB VRAM made it possible to use a batch size of 8192, which stabilized gradient estimates within the one-cycle schedule and ensured efficient utilization of the parallel compute units.

After implementation, the quality of the surrogate model was evaluated using the regression metrics defined in Table~\ref{mathMetricsTable}. The corresponding results are summarized in Table~\ref{tab:surrogate_metrics}. Overall, the metrics show that the proposed residual deep MLP approximates the target quantities of the waveform descriptors with high accuracy and stable generalization across all considered network configurations.
For the single module configuration, both the training and test datasets exhibit median NRMSE values below 1.0\%. The maximum NRMSE values remain low as well, with an upper bound of 1.4\%, indicating that no pronounced local approximation errors occur. The consistently high Pearson correlation coefficients of 0.99 further demonstrate that the internal structure of the waveform representation is preserved reliably in addition to the absolute target values.
A similarly stable behavior is observed for the dual module topology. For both the IC5 and IC7 branches, median NRMSE values of at most 1.1\%, maximum NRMSE values of no more than 2.3\%, and Pearson correlation coefficients of 0.99 are achieved on both training and test data. This confirms that the descriptor spaces associated with each memory module can be reconstructed with high precision independently of one another.
In summary, the implemented surrogate networks provide a reliable deterministic basis for the subsequent classification models.

\begin{table}[ht]
\centering
\caption{Surrogate network regression performance across network configurations (see Figure~\ref{fig:netze})}
\label{tab:surrogate_metrics}
\footnotesize
\renewcommand{\arraystretch}{1.1}
\begin{tabular}{p{1.7cm} p{1.5cm} >{\centering\arraybackslash}p{1.05cm} >{\centering\arraybackslash}p{1.05cm} >{\centering\arraybackslash}p{1.05cm}}
\hline
Type & Configuration & Median NRMSE & Max NRMSE & Median PCC \\
\hline
\multirow{2}{2.1cm}{\textbf{Single-module\\(see Fig.~\ref{fig:netze} (a))}}
 & IC7 - Train & 0.006 & 0.010 & 0.99 \\
 & IC7 - Test  & 0.007 & 0.014 & 0.99 \\
\hline
\multirow{4}{2.1cm}{\textbf{Dual-module\\(see Fig.~\ref{fig:netze} (b))}}
 & IC5 - Train & 0.003 & 0.013 & 0.99 \\
 & IC5 - Test  & 0.010 & 0.018 & 0.99 \\
 & IC7 - Train & 0.003 & 0.004 & 0.99 \\
 & IC7 - Test  & 0.011 & 0.023 & 0.99 \\
\hline
\end{tabular}
\end{table}

\begin{figure}[h]
\centering
    \includegraphics[width=0.48\textwidth]{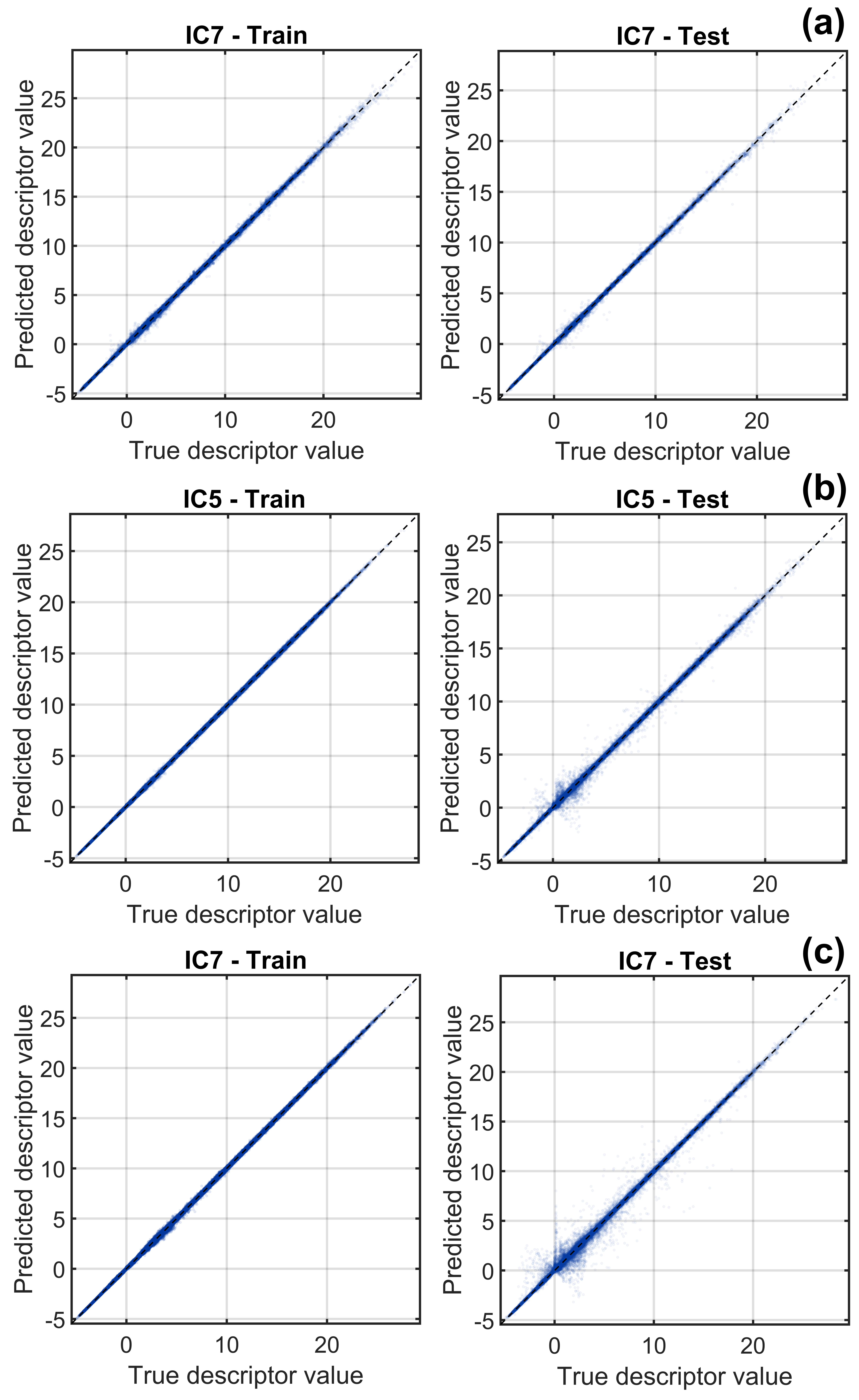}
    \caption{Combined scatter plots of the 12 waveform descriptors for training and test data: (a) Single-module configuration - IC7 (see Fig.~\ref{fig:netze}(a)); (b) Dual-module configuration - IC5 (see Fig.~\ref{fig:netze}(b)); (c) Dual-module configuration - IC7 (see Fig.~\ref{fig:netze}(b))}
    \label{fig:scatterCombWaveform}
\end{figure}

For completeness, Figure \ref{fig:scatterCombWaveform} shows the scatter plots of the predicted waveform descriptors. All diagrams exhibit a very high level of agreement. Neither systematic deviations nor noticeable dispersion patterns are visible. The data points are tightly concentrated along the ideal line, which further confirms the high precision of the predictions across the entire value range.

\subsection{Implementation of the Decision Tree}
To prevent data leakage throughout the pipeline, the same training data split as in the MLP implementation was used. \mbox{Although} the waveforms measured at the inputs of IC5 and IC7 exhibit different configuration-specific signal characteristics, the objective during the training of the decision tree shifts from configuration-dependent signal modeling to a generalized classification of signal quality.
As shown in Figure~\ref{fig:framework}, the decision-tree model operates at the waveform-based AI level. Consequently, the extracted waveform descriptors are merged at the training level. According to Table~\ref{tab:sweeps}, the configuration with two DRAM modules results in a combined training pool of 0.8 × 984,150 waveforms, corresponding to the selected training proportion.
The merging is based exclusively on the training set from the first surrogate-modeling stage, under the assumption that the extracted waveform descriptors represent configuration-independent signal characteristics. The resulting classifier is therefore trained on a shared feature space and is capable of learning generalized decision boundaries for assessing signal quality independently of the underlying network configuration.
The test datasets from IC5 and IC7 remain strictly separated to avoid data leakage and to ensure an unbiased evaluation of generalization capability across both configurations.
The ratio value defined in Equation~\ref{costmat} was set to 8 for both networks in Figure~\ref{fig:netze} to deliberately reduce the class imbalance between minority and majority classes, as described in Section~\ref{decisionTree}.
After implementation, the model quality was evaluated using the metrics presented in Table~\ref{mathMetricsTable}. The results, summarized in Table~\ref{tab:metrics_summary}, show that the decision tree constitutes a well-generalizing model, as the performance indicators between training and test datasets exhibit consistent and balanced behavior for both configurations.
For the single-module case, a sensitivity of at least 99.3\% is achieved, indicating that SI-compliant waveforms are reliably identified as valid. At the same time, a minimum specificity of 99.5\% is obtained, ensuring that non-SI-compliant signals are also correctly detected. A high MCC value of at least 0.95 further confirms that the decision tree makes precise and balanced decisions despite the existing class imbalance.
For the configuration with two DRAM modules, similarly strong results are observed. With a sensitivity of at least 98.4\% and a specificity of at least 98.2\% in the test datasets, high detection rates for both SI-compliant and non-SI-compliant waveforms are achieved. A minimum MCC value above 0.85 demonstrates the reliable performance of the decision tree in the case of a very large number of waveforms.
In summary, the decision tree achieves a balanced trade-off between sensitivity and specificity and exhibits stable generalization capability across different network configurations.

\begin{table}[ht]
\centering
\caption{Decision tree performance across network configurations (see Figure~\ref{fig:netze})}
\label{tab:metrics_summary}
\footnotesize
\renewcommand{\arraystretch}{1.1}
\begin{tabular}{p{1.7cm} p{1.6cm} >{\centering\arraybackslash}p{1.1cm} >{\centering\arraybackslash}p{1cm} >{\centering\arraybackslash}p{1cm}}
\hline
Type & Configuration & Sensitivity & Specificity & MCC \\
\hline

\multirow{2}{2cm}{\textbf{Single-module (see Fig.~\ref{fig:netze} (a))}}
 & IC7 - Train & 0.996 & 0.995 & 0.95 \\
 & IC7 - Test  &  0.993 & 0.995 & 0.95 \\

\hline

\multirow{4}{2cm}{\textbf{Dual-module (see Fig.~\ref{fig:netze} (b))}}
 & \mbox{Train (IC5~+} \mbox{IC7 combined)} & 0.995 & 0.989 & 0.90 \\
 & IC5 - Test & 0.984 & 0.982 & 0.85 \\
 & IC7 - Test & 0.990 & 0.994 & 0.93 \\

\hline
\end{tabular}
\end{table}

\subsection{Implementation of a Multilayer Perceptron Network for Wasserstein Value Prediction}
\begin{figure}[h]
\centering
    \includegraphics[width=0.49\textwidth]{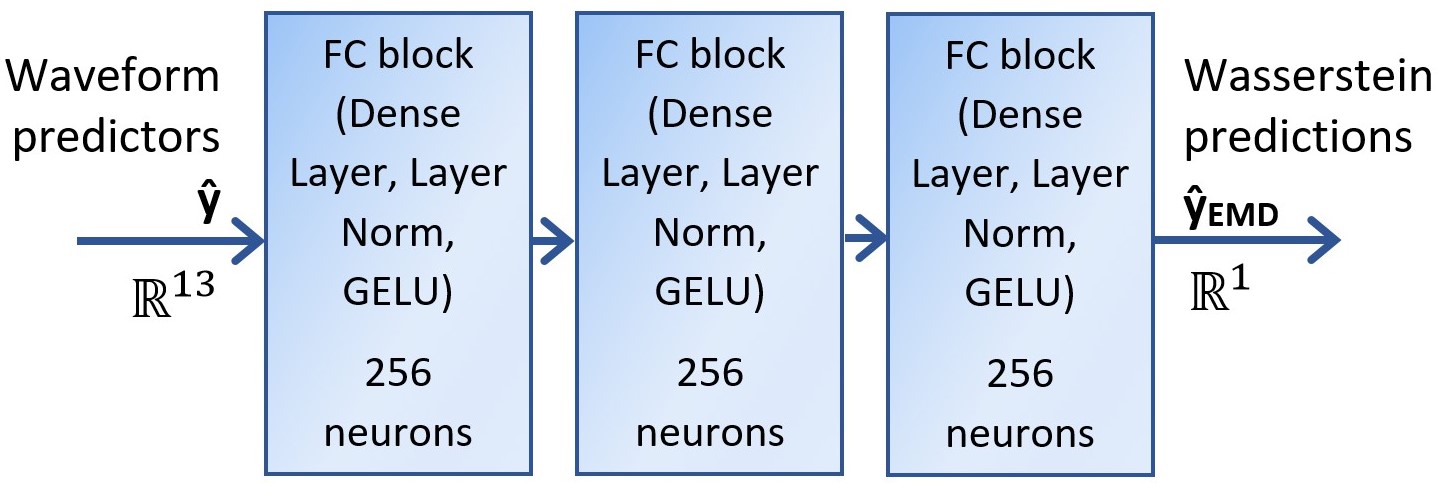}
    \caption{The Wasserstein surrogate model maps the predicted waveform descriptors $\hat{y} \in \mathbb{R}^{12}$ to the estimated Wasserstein distance $\hat{y}_{EMD} \in \mathbb{R}^{1}$ using three sequential fully connected layers}
    \label{fig:confSurrEMD}
\end{figure}

The model implemented for predicting the Wasserstein distance was designed as a sequential deep multilayer perceptron (MLP). In contrast to the high dimensional residual architecture used for the waveform descriptors, a simpler structure was chosen here to account for the lower dimensionality of the target value ($\mathbb{R}^{1}$) and to avoid overfitting. The architecture of the Wasserstein surrogate model is shown in Figure \ref{fig:confSurrEMD}. Within the optimization Framework (see Figure \ref{fig:framework}), the network serves as an accurate estimator of the similarity of the signal distributions (EMD).

The structural design consists of three consecutive fully connected blocks, each comprising 256 neurons. To stabilize the training process and ensure a robust gradient flow, each block applies layer normalization after the dense layer, followed by a nonlinear activation using the GELU function. The final output layer consists of a single neuron with linear activation, which performs the regression of the normalized Wasserstein distance. The Wasserstein surrogate model operates on the same waveform-based AI level as the decision tree and is therefore trained on the combined dataset. The IC5 and IC7 test sets remain separated to ensure an unbiased evaluation of generalization.

Optimization was carried out over 400 epochs using a one cycle learning rate schedule as described in Section \ref{MLPDescr}. After a warm up phase of 10\%, the learning rate increases to a maximum of $1e-3$ (single module) or $2e-3$ (dual module), before decaying to a minimum of $1e-6$ via cosine annealing. In the training runs, the batch size was again set to 8192.

Table \ref{tab:metrics_summaryEMD} summarizes the results, following the structure of Section \ref{MLPDescr}. For the configurations considered, the model achieved NRMSE values below 1\% both during training and on the test data, indicating strong generalization capability without pronounced local approximation errors. In addition to the absolute error values, the Pearson correlation coefficients of 0.99 further confirm that the relationships in the Wasserstein distances are captured accurately.

\begin{table}[ht]\centering
\caption{Performance of the Wasserstein-Surrogate across network configurations (see Figure~\ref{fig:netze})}
\label{tab:metrics_summaryEMD}
\footnotesize
\renewcommand{\arraystretch}{1.1}
\begin{tabular}{p{1.7cm} p{1.6cm} >{\centering\arraybackslash}p{1.1cm} >{\centering\arraybackslash}p{1cm}}
\hline
Type & Configuration & NRMSE & PCC \\
\hline

\multirow{2}{2cm}{\textbf{Single-module (see Fig.~\ref{fig:netze} (a))}}
 & IC7 - Train & 0.002 & 0.99 \\
 & IC7 - Test  &  0.004 & 0.99 \\

\hline

\multirow{4}{2cm}{\textbf{Dual-module (see Fig.~\ref{fig:netze} (b))}}
 & \mbox{Train (IC5~+} \mbox{IC7 combined)} & 0.002 & 0.99 \\
 & IC5 - Test & 0.002 & 0.99 \\
 & IC7 - Test & 0.002 & 0.99 \\

\hline
\end{tabular}
\end{table}

\begin{figure}[h]
\centering
    \includegraphics[width=0.48\textwidth]{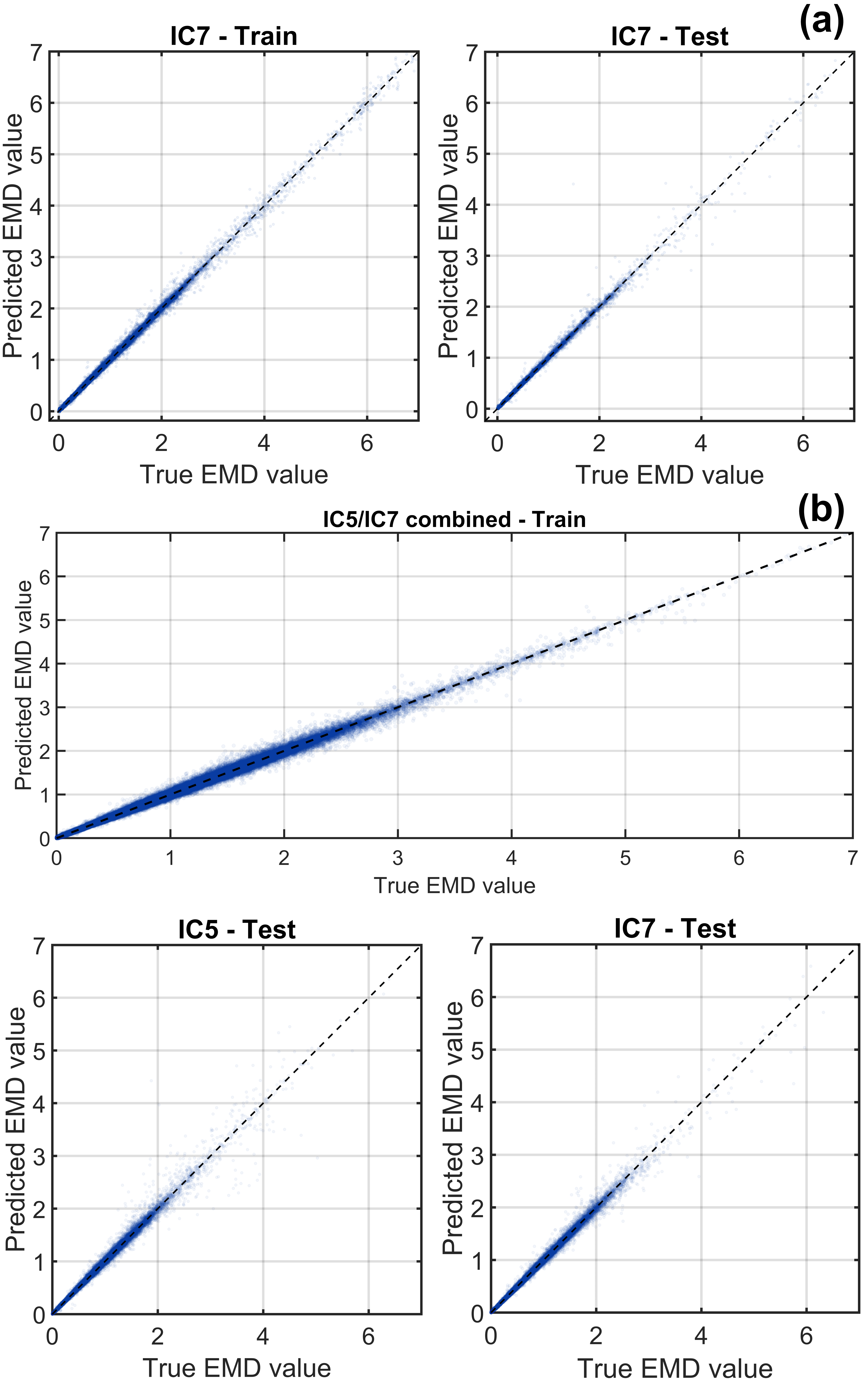}
    \caption{Scatter plots of the EMD values for training and test data: (a) Single-module configuration - IC7 (see Fig.~\ref{fig:netze}(a)); (b) Dual-module configuration - IC5 and IC7 (see Fig.~\ref{fig:netze}(b))}
    \label{fig:wasserstein_scatter}
\end{figure}

The scatter plots in Figure \ref{fig:wasserstein_scatter} demonstrate strong approximation performance across all evaluated configurations, with the data points closely following the ideal line and exhibiting no systematic bias. This confirms that the implemented cascade provides a robust foundation for the subsequent classification of system states. It should be noted that the values in the scatter plot lie on larger scales and are no longer concentrated mainly between 0 and 1. This is due to the fact that a logarithmic transformation followed by inversion and exponentiation of the EMD values was applied. The purpose of this transformation was to map originally small distances, which correspond to high similarity, to larger numerical values and thereby stretch the relevant similarity region. As a result, fine differences between good matches become more visible and receive a stronger weighting during training.

\section{System Integration and Earth Mover’s Distance-Based Evaluation}\label{integration}
After the implementation, the individual models were integrated into a consistent framework. To assess the performance and stability of the complete pipeline, representative evaluations were conducted. Starting from the single-receiver network shown in Figure 1 (a), two central application scenarios are examined in detail in the following. These scenarios are designed to evaluate the predictive capabilities of the framework under varying conditions. The first application scenario primarily serves as a proof of concept, demonstrating that the system is capable of generating valid design proposals within a plausible defined parameter space.

\begin{center}
    \begin{minipage}{\linewidth}
	\raggedright
        \textit{Parameter Space 1 (given by the designer) - single DRAM}\\[2mm]
        \centering
        \resizebox{\linewidth}{!}{
            \begin{tabular}{l c}
            \toprule
            \textbf{Parameter} & \textbf{Range (17,384 combinations)} \\
            
            Trace $L_1$ & \SIrange{20}{125}{\milli\meter} ($\Delta L = \SI{1}{\milli\meter}$) \\
            Trace $L_2$ & \SIrange{2}{5}{\milli\meter} ($\Delta L = \SI{1}{\milli\meter}$) \\
            Resistor (TTL branch) & \SIrange{40}{80}{\ohm} ($\Delta R = \SI{1}{\ohm}$) \\
            \bottomrule
            \end{tabular}
        }
    \end{minipage}
\end{center}

The choice of these bounds follows specific design considerations. The restrictive upper limit for the stub length L2 reflects the fundamental design rule of keeping stubs as short as possible to avoid signal reflections. In contrast, the interval selected for L1 represents typical topological constraints, where variations in trace length are physically limited by layout restrictions. The prediction results for the five best design variants, ranked according to the Earth Mover’s Distance (EMD), are summarized in Table~\ref{tab:emd_ga_comparison}. To validate the performance of the proposed method, an additional reference solution was computed using a Genetic Algorithm (GA). This serves as a benchmark to assess the accuracy of the framework’s predicted optima as well as the required computation time. It can be observed that, due to the iterative nature of the Genetic Algorithm, a suitable solution is obtained only after more than one minute of evolutionary progress. This solution matches the top-5 list generated by the EMD-based approach presented here, confirming the validity and efficiency of the proposed framework.

\begin{table}[h!]
\centering
\caption{Comparison of the top 5 EMD-based parameter solutions and the Genetic Algorithm result for parameter space 1}
\label{tab:emd_ga_comparison}
\renewcommand{\arraystretch}{1.3}
\begin{tabular}{|>{\centering\arraybackslash}m{1.6cm}|
                >{\centering\arraybackslash}m{2.5cm}|
                >{\centering\arraybackslash}m{2cm}|}
\hline
\textbf{Method} & \textbf{Best Parameter Combination(s)} & \textbf{Computation Time [s]} \\
\hline
Top 5 EMD-based solutions &
\begin{tabular}[c]{@{}c@{}}
$[20,\,2,\,40]$ \\
$[20,\,2,\,41]$ \\
$[20,\,2,\,42]$ \\
$[21,\,2,\,40]$ \\
$[21,\,2,\,41]$
\end{tabular}
& 0.43 \\
\hline
Genetic Algorithm &
$[20,\,2,\,41]$
& 78.04 \\
\hline
\end{tabular}
\end{table}

\begin{figure}[h]
\centering
    \includegraphics[width=0.48\textwidth]{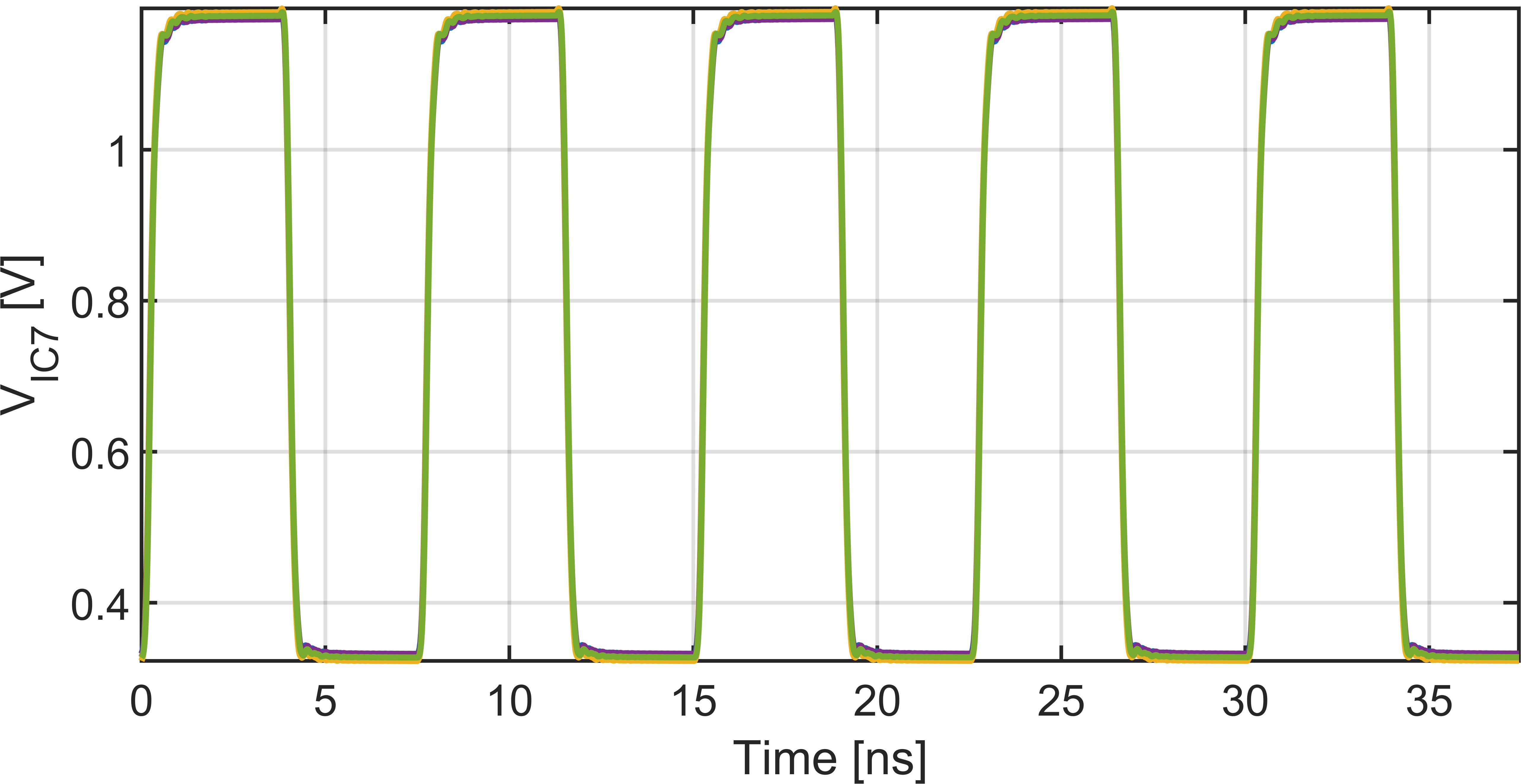}
    \caption{Re-simulation of the top-5 parameter combinations proposed in Table \ref{tab:emd_ga_comparison}}
    \label{fig:paramSpace1}
\end{figure}

In Figure \ref{fig:paramSpace1}, all design proposals generated by the EMD-based method are shown in the form of corresponding simulation results. Each waveform was re-simulated based on the design parameters proposed by the framework, allowing a direct assessment of their signal-integrity performance. It can be observed that all resulting waveforms exhibit an adequate rectangular shape and, in the high-state plateau, no overshoot occurs. This is achieved under the constraint that the signal rise time is maximized.

\begin{figure}[h]
\centering
    \includegraphics[width=0.48\textwidth]{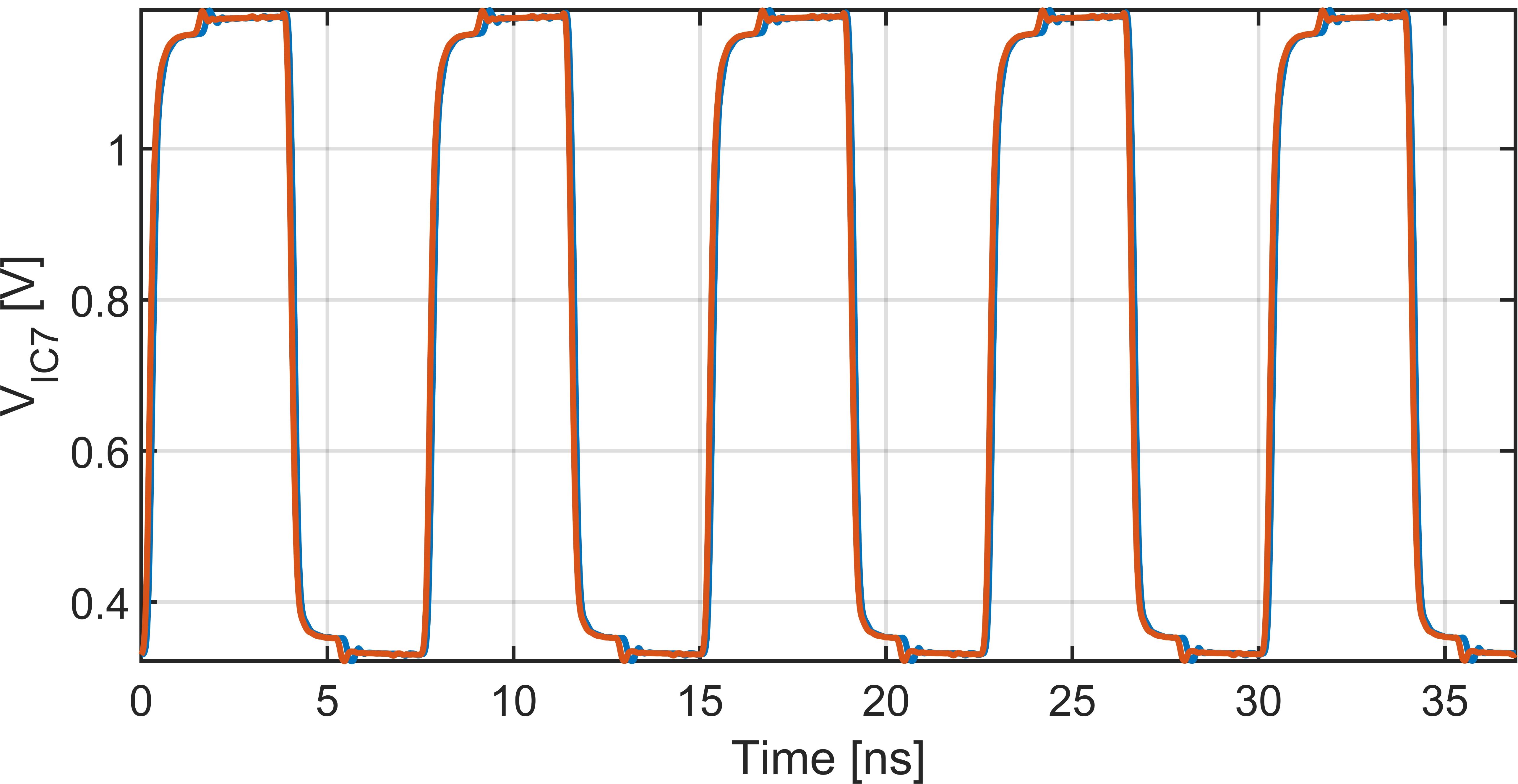}
    \caption{Re-simulation of lower-ranked parameter combinations for parameter space 1 (single DRAM module)}
    \label{fig:paramSpace1lower}
\end{figure}

Two lower-ranked parameter combinations from the EMD-sorted list of SI-compliant signals within parameter space 1 were also re-simulated, namely the sets [121, 4, 41] and [110, 2, 41]. The corresponding results are shown in Figure~\ref{fig:paramSpace1lower}. It can be clearly observed that these signals still exhibit an adequate waveform with an acceptable rise time. However, compared to the top-5 designs shown in Figure~\ref{fig:paramSpace1}, their waveforms display a more irregular shape in the high-state plateau. This highlights the advantage of selecting the best parameter sets using the Earth Mover's Distance, in contrast to the statistical method presented in \cite{ecik2024}. After narrowing down the desired or relevant parameter ranges, the designer can use the EMD-based ranking to perform a more differentiated selection among the SI-compliant candidates.

For an additional test, a second parameter space was defined, in which the selected value ranges were intentionally chosen far outside typical and reasonable design rules.

\begin{center}
    \begin{minipage}{\linewidth}
        \raggedright 
        \textit{Parameter Space 2 (given by the designer) - single DRAM}\\[2mm]
        
        \centering 
        \resizebox{\linewidth}{!}{%
            \begin{tabular}{l c}
            \toprule
            \textbf{Parameter} & \textbf{Range (396 combinations)} \\
            \midrule
            Trace $L_1$ & \SIrange{150}{155}{\milli\meter} ($\Delta L = \SI{1}{\milli\meter}$) \\
            Trace $L_2$ & \SIrange{120}{125}{\milli\meter} ($\Delta L = \SI{1}{\milli\meter}$) \\
            Resistor (TTL branch) & \SIrange{150}{160}{\ohm} ($\Delta R = \SI{1}{\ohm}$) \\
            \bottomrule
            \end{tabular}
        }
    \end{minipage}
\end{center}

\begin{figure}[h]
\centering
    \includegraphics[width=0.48\textwidth]{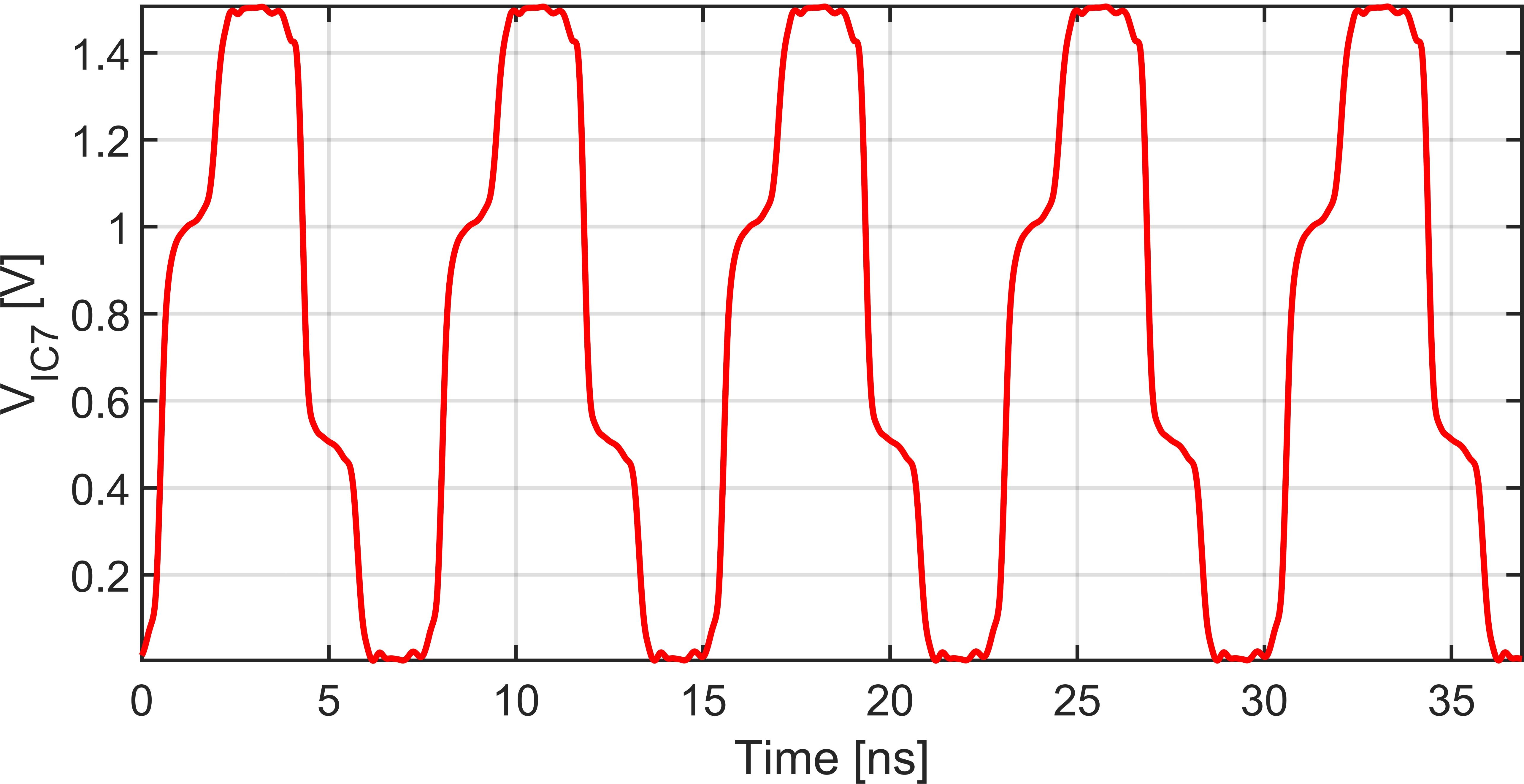}
    \caption{Re-simulation of the parameter combination [150, 120, 150] proposed by the GA (parameter space 2; single DRAM module)}
    \label{fig:paramSpace1nonSI}
\end{figure}

For this additional experiment, the implemented framework was able to determine within only 0.29 seconds that no SI-compliant solutions existed in the selected parameter space 2. In contrast, the Genetic Algorithm proposed the parameter combination [150, 120, 150] in 3.5 seconds. As shown in Figure \ref{fig:paramSpace1nonSI}, this result is clearly not SI-compliant. The waveform does not meet the required rise-time characteristics due to the lack of monotonicity in the rising edge. In this case, the Genetic Algorithm simply operated within the specified parameter bounds and returned the best candidate according to its internal optimization criteria, even though the resulting waveform is unsuitable.

This example highlights the advantage of the proposed decision-tree-based approach, which performs an internal SI check and immediately warns the designer when no SI-compliant signals are present in the selected parameter region. While it would be possible to extend the Genetic Algorithm with additional hand-crafted SI constraints to avoid such cases, doing so would weaken the goal of a generalizable, model-driven approach and would move the method closer to the rigid rule-based structures discussed before.

Analogous to the procedure described above, two parameter ranges were also investigated for the network with two memory modules (see Figure\ref{fig:netze}~(b)). 

\begin{center}
    \begin{minipage}{\linewidth}
        \raggedright 
        \textit{Parameter Space 1 (given by the designer) - dual DRAM}\\[2mm]
       
        \centering 
        \resizebox{\linewidth}{!}{
            \begin{tabular}{l c}
            \toprule
            \textbf{Parameter} & \textbf{Range (19,844 combinations)} \\
            \midrule
            Trace $L_1$ & \SIrange{10}{20}{\milli\meter} ($\Delta L = \SI{1}{\milli\meter}$) \\
            Trace $L_2$ & \SIrange{10}{20}{\milli\meter} ($\Delta L = \SI{1}{\milli\meter}$) \\
           Trace $L_3$ & \SIrange{2}{5}{\milli\meter} ($\Delta L = \SI{1}{\milli\meter}$) \\
            Resistor (TTL branch) & \SIrange{30}{70}{\ohm} ($\Delta R = \SI{1}{\ohm}$) \\
            \bottomrule
            \end{tabular}
        }
    \end{minipage}
\end{center}

\begin{center}
    \begin{minipage}{\linewidth}
        \raggedright 
        \textit{Parameter Space 2 (given by the designer) - dual DRAM}\\[2mm]
        
        \centering 
        \resizebox{\linewidth}{!}{%
            \begin{tabular}{l c}
            \toprule
            \textbf{Parameter} & \textbf{Range (2,376 combinations)} \\
            \midrule
            Trace $L_1$ & \SIrange{150}{155}{\milli\meter} ($\Delta L = \SI{1}{\milli\meter}$) \\
            Trace $L_2$ & \SIrange{120}{125}{\milli\meter} ($\Delta L = \SI{1}{\milli\meter}$) \\
           Trace $L_3$ & \SIrange{100}{105}{\milli\meter} ($\Delta L = \SI{1}{\milli\meter}$) \\
            Resistor (TTL branch) & \SIrange{150}{160}{\ohm} ($\Delta R = \SI{1}{\ohm}$) \\
            \bottomrule
            \end{tabular}
        }
    \end{minipage}
\end{center}

\begin{figure}[h]
\centering
    \includegraphics[width=0.48\textwidth]{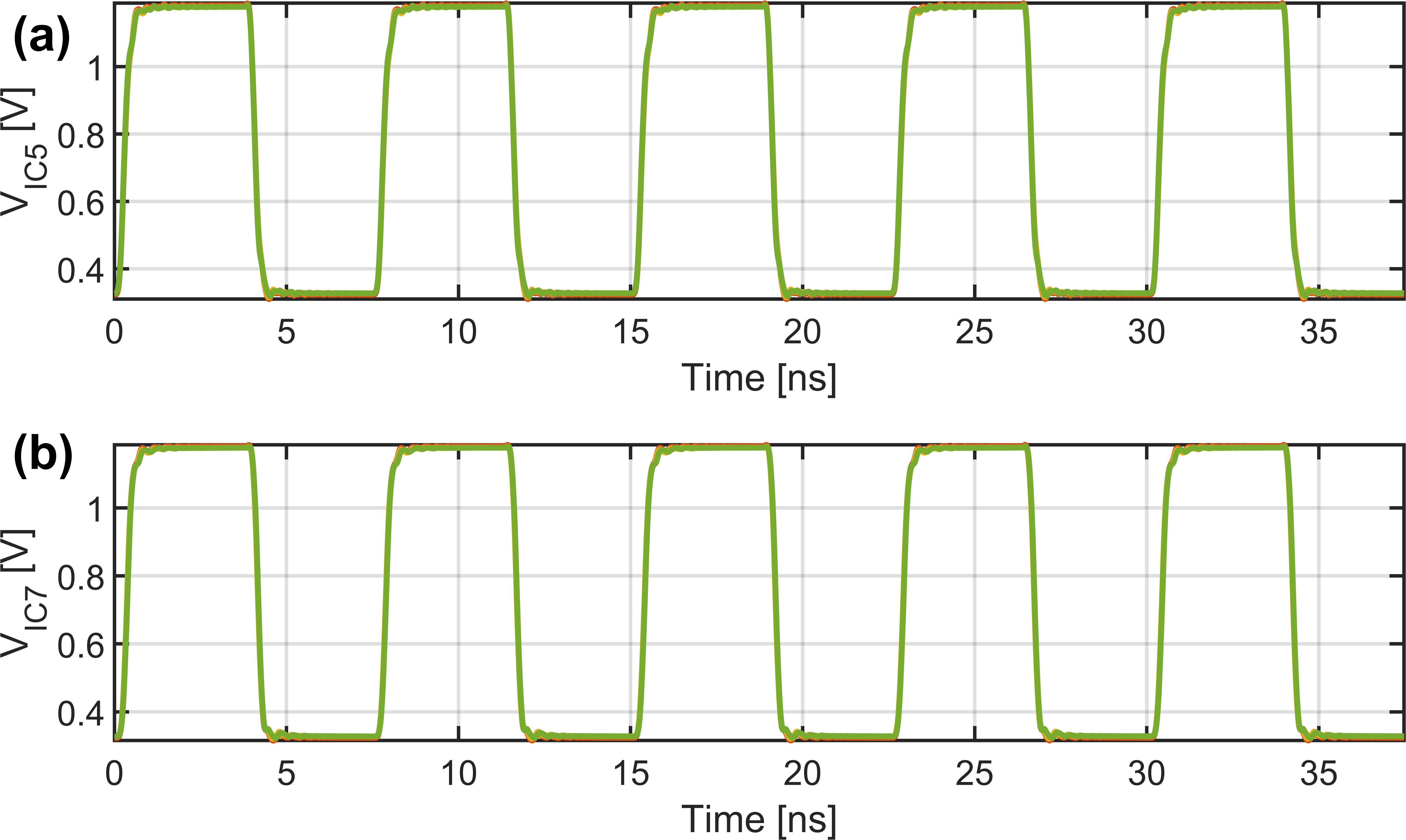}
    \caption{Re-simulation of the top-5 parameter combinations (parameter space 1; dual DRAM configuration): (a) IC5 SI-compliant waveforms; (b) IC7 SI-compliant waveforms}
    \label{fig:paramSpace1dual}
\end{figure}

\begin{figure}[h]
\centering
    \includegraphics[width=0.48\textwidth]{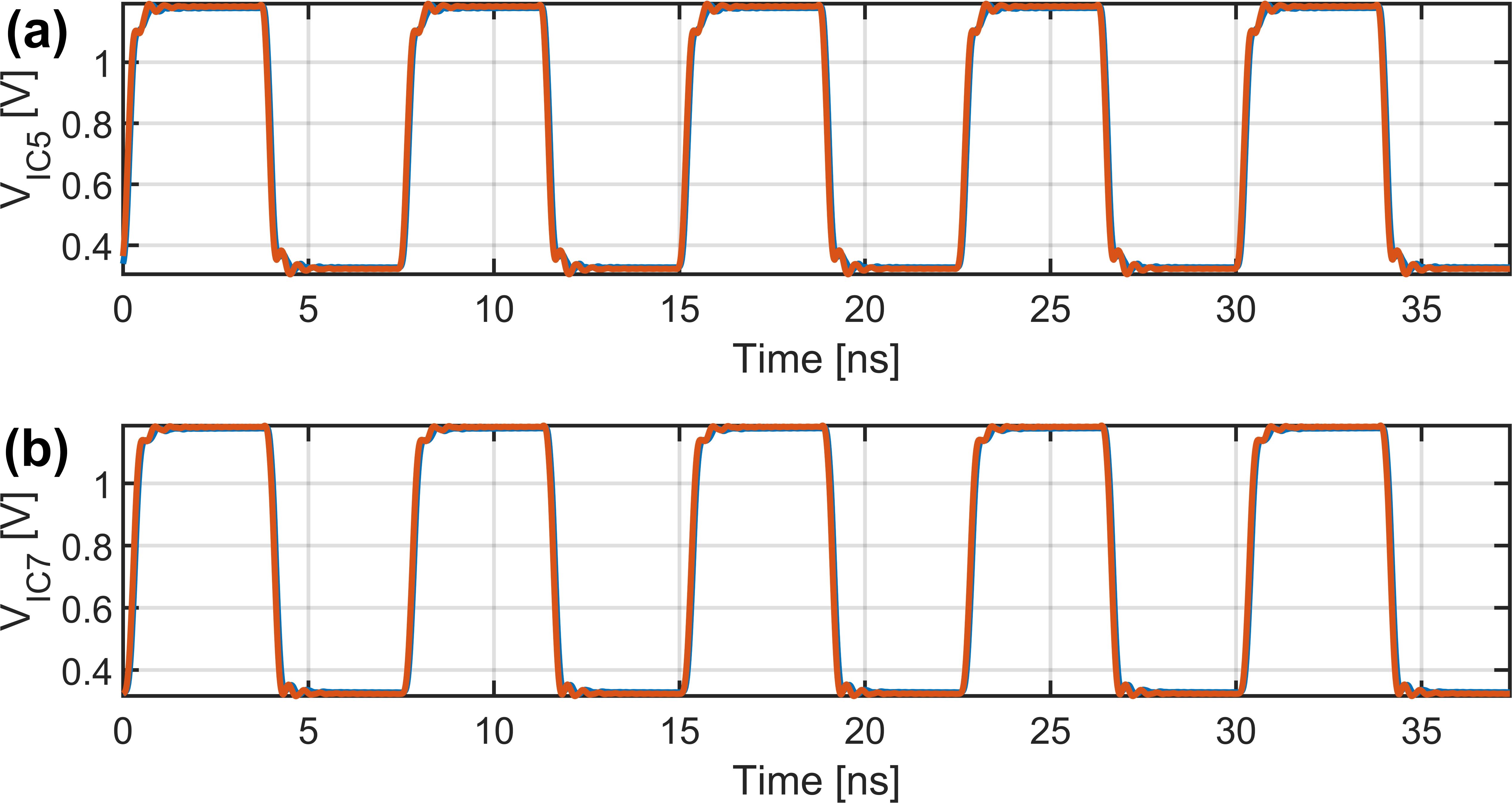}
    \caption{Re-simulation of lower-ranked parameter combinations ([19,\,17,\,4,\,41], [13,\,18,\,5,\,42]) from EMD-sorted list (parameter space 1; dual DRAM configuration): (a) IC5 SI-compliant waveforms; (b) IC7 SI-compliant waveforms}
    \label{fig:paramSpace1lowerdual}
\end{figure}

\begin{figure}[h]
\centering
    \includegraphics[width=0.48\textwidth]{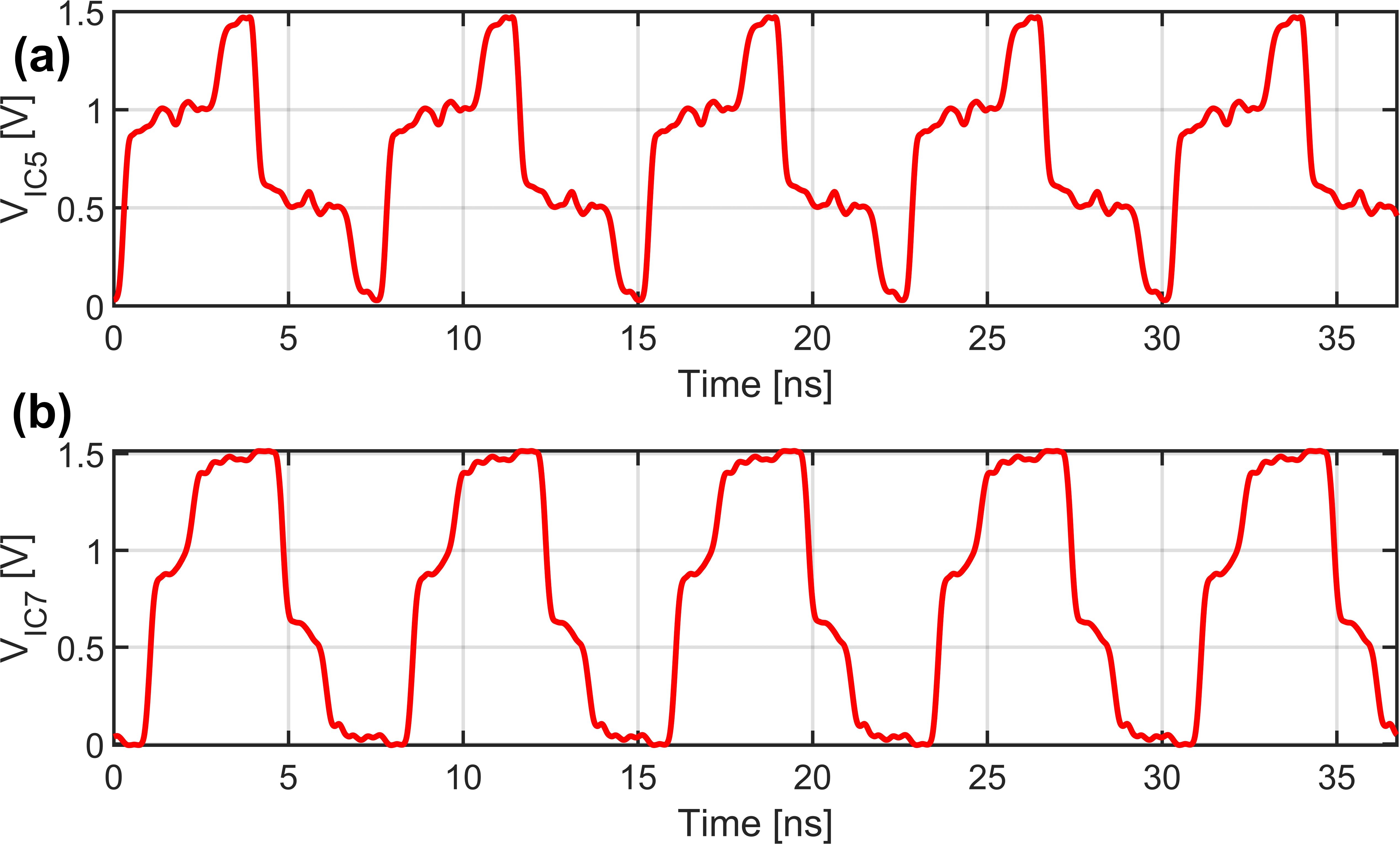}
    \caption{Re-simulation of the parameter combination [150, 120, 100, 152] proposed by the GA (parameter space 2; dual DRAM configuration): (a) IC5 waveforms; (b) IC7 waveforms}
    \label{fig:paramSpace1nonSIdual}
\end{figure}

The evaluation and re-simulation results are summarized in Table \ref{tab:emd_ga_comparisondual} and figures \ref{fig:paramSpace1dual} to \ref{fig:paramSpace1nonSIdual}. The presentation follows the same logic as applied to the single memory module network (see Figure \ref{fig:netze}~(a)).

\begin{table}[h!]
\centering
\caption{Comparison of the solutions for parameter space 1 and 2 (dual DRAM configuration)}
\label{tab:emd_ga_comparisondual}
\renewcommand{\arraystretch}{1.2}
\begin{tabular}{|>{\centering\arraybackslash}m{2.7cm}|
                >{\centering\arraybackslash}m{2.5cm}|
                >{\centering\arraybackslash}m{2cm}|}
\hline
\textbf{Method} & \textbf{Best Parameter Combination(s)} & \textbf{Computation Time [s]} \\
\hline
Top 5 EMD-based solutions (parameter space 1) &
\begin{tabular}[c]{@{}c@{}}
$[20,\,10,\,4,\,41]$ \\
$[20,\,10,\,5,\,42]$ \\
$[20,\,10,\,5,\,41]$ \\
$[20,\,11,\,5,\,41]$ \\
$[20,\,11,\,4,\,41]$
\end{tabular}
& 0.89 \\
\hline
Genetic Algorithm (parameter space 1) &
$[20,\,10,\,4,\,41]$
& 60.34 \\
\hline
Top 5 EMD-based solutions (parameter space 2) &
No SI-compliant solutions found
& 0.27 \\
\hline
Genetic Algorithm (parameter space 2) &
$[150, 120, 100, 152]$
& 3.22 \\
\hline
\end{tabular}
\end{table}

\conclusions 
The results presented in Chapter \ref{integration} demonstrate the efficiency of the chosen approach. By combining neural surrogate models, a decision tree, and the Earth Mover's Distance, SI-compliant signals could be evaluated with high accuracy, and valid design regions were identified deterministically. The proposed Framework, validated through re-simulations, enables a physically motivated and granular ranking of solutions relative to an ideal reference signal via the EMD. This provides a more differentiated analysis than the purely statistical evaluation in \cite{ecik2024} and offers designers the flexibility to investigate specific parameter regions in a targeted manner. Due to its low computational cost, this efficient approach represents a powerful alternative to classical optimization strategies.

Looking ahead, the method should be extended and transferred to technologies such as DDR4 and DDR5. These standards introduce specific challenges, including substantially higher data rates and the adaptive adjustment of on-chip reference voltages. Since evaluating SI compliance based on conventional transient waveforms becomes increasingly impractical for DDR4/5, the identification of valid parameter configurations must shift toward alternative target metrics such as eye diagrams and bit-error rates.

\codedataavailability{Due to project restrictions, the source codes and datasets are not publicly available. For more information, please contact the main author.} 

\authorcontribution{EE designed the AI approaches and implemented the resulting models based on decision trees, including the corresponding training procedures. In addition, EE developed the concept of combining the Earth Mover's Distance with further AI models to enable a more precise identification of SI-compliant signals. EE also carried out essential adaptations and further developments of existing approaches as well as code components from earlier work. Moreover, EE authored both the initial and the final version of the manuscript. WJ and RB contributed to a deeper understanding of the SI knowledge domain and to the evaluation of the prediction results. They also supported the further development of the AI models in the context of SI-specific design requirements and participated in revising the manuscript. JW and EE incorporated results from earlier work on the use of genetic algorithms into this study in a revised form. JW additionally discussed the results together with WJ, JG, and RB and contributed editorial input. JG supervised the entire research process at the Information Processing Lab and performed the final review.
} 

\competinginterests{The contact author has declared that none of the authors has any competing interests.} 

\begin{acknowledgements}
This work is funded as part of the research project KI4BoardNet in the funding programme MANNHEIM (BMBF) (Grant numbers 16ME0777/9). The responsibility for this publication is held by the authors only. The original presentation was held at the U.R.S.I. Kleinheubach Conference 2025. 
\end{acknowledgements}

\FloatBarrier
\appendix
\section{Waveform Descriptors for the Training of the AI-Models}    
\label{appa}
A total of 12 waveform descriptors were extracted from each signal for training the AI models. Table \ref{tab:featTab} provides an overview of all computed descriptors.

\begin{table}[h]
\centering
\caption{Overview of extracted waveform descriptors}
\label{tab:featTab}
\footnotesize
\begin{tabular}{>{\raggedright\arraybackslash}p{1.8cm} p{2.5cm} >{\centering\arraybackslash}p{1.1cm} c}
\hline
Descriptor & Description & Scope & Normalized \\
\hline
Upper edge slope & Slope to 90\% peak from mid-level baseline & windowed & No \\
Maximum slope & Maximum of smoothed signal slope & windowed & No \\
Mean squared difference & Mean squared difference of successive samples & windowed & No \\
Skewness (norm.) & Skewness of normalized window & windowed & Yes \\
Kurtosis (norm.) & Kurtosis of normalized window & windowed & Yes \\
Skewness (glob.) & Global skewness & global & No \\
Kurtosis (glob.) & Global kurtosis & global & No \\
Peak-to-peak amplitude & Maximum minus minimum of signal & global & No \\
Energy variance & Variance of the energy distribution & windowed & No \\
KDE entropy & Entropy via Kernel Density Estimation & windowed & Yes \\
Wavelet entropy & Wavelet entropy (1st-order Haar) & windowed & Yes \\
95\% Z-Score & Statistical distance of 95th percentile & windowed & Yes \\
\hline
\end{tabular}
\end{table}

For completeness, the mathematical definitions of the calculated descriptor set are provided below. The formulations are based on established definitions in \citet{mishra}, \citet{peeters}, \citet{blanco}, \citet{shannon}, \citet{silverm} and \citet{matlab}. The same formulations are used for all signals, with differences only in the input domain: descriptors are computed either over the entire signal (global descriptors) or within a detected signal segment (windowed descriptors). In the following equations, $x_i$ and $w_i$ denote the discrete amplitude and energy samples of the respective signal or segment, where $i = 1, \dots, N$ and $N$ represents the total number of samples. The signal window is defined based on a thresholding approach relative to the mean signal level. The onset of the window corresponds to the first sample exceeding this threshold and is used as the reference point for subsequent feature extraction. Here, the mid-level baseline corresponds to the first value of the detected signal window and serves as the reference point for feature computation.

The waveform descriptors are categorized into time-domain and time-frequency-domain descriptors. Entropy-based descriptors are computed across multiple signal representations, including the time and time-frequency domains.

\subsection{Time-domain features}
Time-domain descriptors are computed directly from the raw signal or signal segments. This group includes statistical descriptors, energy-based measures and classical temporal characteristics.

The mean and standard deviation are used as auxiliary statistical measures and are defined as:
\begin{equation}
\mu = \frac{1}{N} \sum_{i=1}^{N} x_i, \quad
\sigma = \sqrt{\frac{1}{N} \sum_{i=1}^{N} (x_i - \mu)^2}.
\end{equation}

Skewness and kurtosis are computed as standardized central moments:
\begin{align}
& \text{Skewness} = \frac{1}{N} \sum_{i=1}^{N} \left(\frac{x_i - \mu}{\sigma}\right)^3, \\
& \text{Kurtosis} = \frac{1}{N} \sum_{i=1}^{N} \left(\frac{x_i - \mu}{\sigma}\right)^4.
\end{align}

For normalized descriptors, min–max scaling is applied prior to computation as follows:
\begin{equation}
x_{i,\text{norm}} = \frac{x_i - \min(x_i)}{\max(x_i) - \min(x_i)}.
\end{equation}

Energy is defined via squared signal amplitudes:
\begin{equation}
w_i = x_i^2.
\end{equation}
The energy-based descriptor used in this work is the variance of the energy distribution derived from the sequence $w_i$.

Furthermore, the peak-to-peak amplitude is defined as:
\begin{equation}
A_{pp} = \max(x_i) - \min(x_i).
\end{equation}

The upper edge slope is computed as the slope between the signal value at the detected window onset and the first occurrence of 90\% of the maximum signal amplitude within the same window:
\[
\text{UpperEdgeSlope} = \frac{x(t_{90}) - x_{mid-level}}{t_{90} - t_{mid-level}}.
\]

The maximum slope is defined as:
\begin{align}
&m_i = \text{movingAverage}(x_i - x_{i-1}), \\
& \text{MaxSlope} = \max(m_i).
\end{align}

In this work, the mean squared difference is utilized to characterize overshoot behavior. By calculating the average of squared first-order differences, it captures high-frequency variations typical for ringing:
\begin{equation}
\text{MeanSqDiff} = \frac{1}{N-1} \sum_{i=1}^{N-1} (x_{i+1} - x_i)^2.
\end{equation}

The 95\% Z-Score provides a measure of extreme values relative to the distribution:
\begin{equation}
Z_{95} = \frac{P_{95} - \mu}{\sigma}.
\end{equation}

\subsection{Entropy-based features}
All entropy measures are based on the Shannon entropy definition:
\begin{equation}
H = - \sum_i p_i \log_2(p_i),
\label{shannon}
\end{equation}

where \(p_i\) denotes a normalized probability distribution derived from the respective signal representation.

The KDE-based entropy employs a Gaussian Kernel Density Estimation (KDE) to estimate the probability distribution $p_i$ over a fixed grid of $M$ amplitude points $g_i$:
\begin{equation}
p_i = \frac{1}{N \cdot h} \sum_{j=1}^{N} K\left(\frac{g_i - x_j}{h}\right),
\end{equation}
where $K$ denotes the Gaussian kernel, $h$ is the bandwidth, and $x_j$ represents the signal samples. The smoothing factor $h$ controls the trade-off between detail and noise suppression. The resulting discrete distribution $p_i$ is normalized such that $\sum p_i = 1$ and subsequently used to calculate the Shannon entropy according to Equation (\ref{shannon}).

The wavelet entropy used in this work is based on first-order local differences rather than a full multiscale decomposition.

A discrete Haar wavelet is applied at the first decomposition level (level-1 detail coefficients):
\begin{equation}
h = \frac{1}{\sqrt{2}} [1, -1].
\end{equation}

The corresponding wavelet coefficients are obtained via convolution $c_k = (x * h)_k$. Subsequently, an energy representation is constructed from the squared coefficients $E_k = c_k^2$. A normalized probability distribution is then defined as:
\begin{equation}
p_i = \frac{E_k}{\sum_j E_j}.
\end{equation}

Finally, the wavelet entropy is computed using the Shannon entropy definition in Equation (\ref{shannon}), where $p_i$ represents the probability distribution over the respective representation indices.

\section{Physics-Inspired Feature Engineering}\label{appb}
The feature-engineering procedure is defined with respect to the underlying DDR3 fly-by network topology shown in Figure~\ref{fig:netze}. Depending on the configuration considered, either a single or a dual DRAM module setup is used, resulting in a variable number of transmission line parameters. This leads to input sets with a cardinality of two or three line parameters, corresponding to the respective network topology.

\renewcommand{\arraystretch}{1.1}
\begin{table}[ht]
\centering
\caption{Feature set of physics-inspired engineered inputs}
\label{tab:engineered_features}
\footnotesize
\begin{tabular}{>{\raggedright\arraybackslash}p{1.7cm} p{2.1cm} p{3.65cm}}
\hline
Feature & Formula & Physics-inspired Motivation \\
\hline
Normalized base features 
&{\large $ \text{\normalsize{param}}_i = \frac{\text{param}_{\textrm{raw}} - \text{param}_{\textrm{min}}}{\text{param}_{\textrm{max}} - \text{param}_{\textrm{min}}} $}
& Normalized physical input parameters (line lengths, resistance); basis for all derived features \\
Line lengths $L_i$
& $ L_i = \text{param}_{i,\mathrm{Tline}} $
& Geometric line lengths influencing Delay \\
Line length differences
& $ d_{ij} = L_i - L_j $
& Relative line lengths; models propagation delay differences \\
Line termination
& $ R = \text{param}_{\textrm{raw,R}} $
& Termination resistance; affects reflection behavior \\
Reflection factor
& $ \Gamma = \frac{50 - R}{50 + R} $
& Reflection coefficient relative to fixed characteristic impedance of Z=$50\,\Omega$\\
inv(R), abs($\Gamma$), $\Gamma^2$
& $ 1/R,\ |\!\Gamma\!|,\ \Gamma^2 $
& Measures for conductance, reflection magnitude and reflected power \\
Fourier features (1st–4th order)
& $ \Phi_i = \sin(k 2\pi\,\text{param}_i)$, $\cos(k 2\pi\,\text{param}_i)$, $k=1..4 $
& Capture periodic patterns; harmonics up to 4th order \\
Interactions
& $ \text{param}_i \cdot \text{param}_j $
& Nonlinear coupling between parameters \\
Hierarchical interactions
& $ \phi_i \cdot \text{param}_j $
& Fourier × parameter; captures modulated nonlinear relationships \\
Interference features
& $ \sin(2\pi(\text{param}_i \pm \text{param}_j)) $
& Sensitive to relative differences between parameters \\
Quadratic features
& $ \text{param}_i^2 $
& Models nonlinear effects \\
Cubic features
& $ \text{param}_i^3 $
& Captures higher-order nonlinearities \\
Absolute distances
& $ |\text{param}_i - \text{param}_j| $
& Models relative differences and delay variations \\
Quadratic distances
& $ (\text{param}_i - \text{param}_j)^2 $
& Emphasizes larger deviations \\
log(R)
& $ \log(R) $
& Stabilizes large dynamic ranges \\
Log deviation from Z=$50\,\Omega$
& $ \log(|R - 50| + 1) $
& Robust measure of impedance mismatch \\
Rnorm
& {\large $ \frac{R - 50}{50} $}
& Relative deviation from characteristic impedance Z=$50\,\Omega$\\
Rdev$^2$
&$ (R - 50)^2 $
& Quadratic deviation from characteristic impedance Z=$50\,\Omega$\\
Resistive cross features
& $ R \cdot \text{param}_i $
& Coupling between resistance and parameter \\
Reflective cross features
& $ \Gamma \cdot \text{param}_i $
& Coupling between reflection and parameter \\
Quadratic parameters
& $ \sum_i \text{param}_i^2 $
& Global measure of parameter intensity \\
\hline
\end{tabular}
\vspace{-10mm}
\end{table}
\renewcommand{\arraystretch}{1.0}

\section{Mathematical Definintion of the Evaluation Metrics}\label{mathMetrics}
Table \ref{mathMetricsTable} provides an overview of all performance measures applied in this work, together with their mathematical formulations. Because the analysis includes both binary classification and continuous regression tasks, the chosen metrics cover evaluation criteria relevant to both domains.

\renewcommand{\arraystretch}{1.3}
\begin{table}[h]
\centering
\caption{Overview of all evaluation metrics used for model assessment.}
\label{mathMetricsTable}
\begin{tabular}{>{\raggedright\arraybackslash}m{2.05cm} >{\centering\arraybackslash}m{5.6cm}}
\hline
Metric & Mathematical Formulation \\
\hline

Sensitivity \citep[Recall;][]{murphy}
& {\large$\dfrac{TP}{TP+FN}$} \\

Specificity \citep{murphy}
& {\large$\dfrac{TN}{TN+FP}$} \\

MCC \citep{matt}
&\hspace{-4.5mm}{\fontsize{7.5pt}{8pt}\selectfont $\dfrac{TP \cdot TN - FP \cdot FN}
{\sqrt{(TP+FP)(TP+FN)(TN+FP)(TN+FN)}}$} \\

NRMSE \citep[range-normalized;][]{khosh}
& {\large $\dfrac{\sqrt{\frac{1}{N}\sum_{i=1}^{N}(y_i-\hat{y}_i)^2}}
{y_{\max}^{\mathrm{train}} - y_{\min}^{\mathrm{train}}}$} \\

Pearson correlation coefficient $r$ \citep{pearson, murphy}
& {\large $\dfrac{\sum_{i=1}^{N}(y_i-\bar{y})(\hat{y}_i-\bar{\hat{y}})}
{\sqrt{\sum_{i=1}^{N}(y_i-\bar{y})^2
\sum_{i=1}^{N}(\hat{y}_i-\bar{\hat{y}})^2}}$} \\
\hline
\end{tabular}
\smallskip
\begin{minipage}{8.1cm}
   \footnotesize
    \textit{Note:} $y_i$ and $\hat{y}_i$ denote the observed and predicted values for the $i$-th sample; $\bar{y}$ and $\bar{\hat{y}}$ represent their respective means; $N$ is the total number of samples.
\end{minipage}
\end{table}
\renewcommand{\arraystretch}{1.0}

\noappendix      

\appendixfigures 

\appendixtables 

\FloatBarrier

\end{document}